\documentclass[12pt]{iopart}
\usepackage{iopams}
\usepackage{graphicx}
  
\begin{document}

\title[Force of light on a two-level atom near an ultrathin optical fiber]{Force of light on a two-level atom near an ultrathin optical fiber}

\author{Fam Le Kien$^1$, D F Kornovan$^{2,3}$,  S Sahar S Hejazi$^1$, Viet Giang Truong$^4$, M I Petrov$^{2,3}$, S\'{i}le Nic Chormaic$^{4}$ and Thomas Busch$^1$}

\address{$^1$Quantum Systems Unit, Okinawa Institute of Science and Technology Graduate University, Onna, Okinawa 904-0495, Japan}
\address{$^2$ ITMO University, Birzhevaya liniya 14, 199034 St. Petersburg, Russia}
\address{$^3$ St. Petersburg Academic University, 8/3 Khlopina str., 194021 St. Petersburg, Russia}
\address{$^4$Light-Matter Interactions Unit, Okinawa Institute of Science and Technology Graduate University, Onna, Okinawa 904-0495, Japan}

\ead{kienle.pham@oist.jp}
\vspace{10pt}
\begin{indented}
\item[]\today
\end{indented}

\begin{abstract}
We study the force of light on a two-level atom near an ultrathin optical fiber
using the mode function method and the Green tensor technique. We show that the total force consists of the driving-field force, the spontaneous-emission recoil force, and the fiber-induced van der Waals potential force. 
Due to the existence of a nonzero axial component of the field in a guided mode, 
the Rabi frequency and, hence, the magnitude of the force of the guided driving field may depend on the propagation direction.  
When the atomic dipole rotates in the meridional plane, the spontaneous-emission recoil force may arise as a result of the asymmetric spontaneous emission with respect to opposite propagation directions.
The van der Waals potential for the atom in the ground state is off-resonant and opposite to the off-resonant part of the van der Waals potential for the atom in the excited state. Unlike the potential for the ground state, the potential for the excited state may oscillate depending on the distance from the atom to the fiber surface. 
\end{abstract}

\vspace{2pc}
\maketitle

\section{Introduction}
\label{sec:introduction}

It is known that the interaction between light and an atom leads to an optical force.
Exerting controllable optical forces on atoms finds important applications in many areas of physics, in particular in laser cooling and trapping. 
A large number of schemes for such phenomena have been proposed, studied, and implemented \cite{coolingbook,dipole force}.
A common feature of the cooling and trapping schemes for atoms in free space is that 
the average of the recoil over many spontaneous emission events results in a zero net effect on the momentum transfer.
Thence, the optical forces on atoms in free space are determined by only the absorption and stimulated emission of light and the light shifts of the ground and excited states \cite{coolingbook,dipole force}.

An atom near a material object undergoes a dispersion force, which can be called the van der Waals force or the Casimir-Polder force in the nonretarded or retarded interaction regime \cite{Hoinkes, Klimchitskaya2009, Buhmann book I, Buhmann book II}. 
The van der Waals interactions between atoms and cylinders have been studied \cite{Mehl1980,Toigo1982,Boustimi2002,Boustimi2003,Eberlein2007,Eberlein2009,Minogin2010,Minogin2012}. 
In most of the previous work, the atoms were considered as point-like polarizable particles.
When an atom is driven by an external field near an object, the van der Waals interaction depends on the atomic excitation.
In addition, the atom undergoes a radiation force, which depends on the field intensity, the field polarization, and the atomic dipole orientation. Moreover, due to the presence of the object, a nonzero spontaneous emission recoil force may appear.

Indeed, for atoms near a nanofiber \cite{Fam2014,Petersen2014,Mitsch14b,sponhigh,Scheel2015,Jacob2016}, 
a flat surface \cite{Jacob2016,flat,Buhmann2018}, a photonic topological material \cite{Anzetta2018a,Anzetta2018b}, 
a photonic crystal waveguide \cite{leFeberScience2015}, or a nonreciprocal medium \cite{Fuchs2017}, spontaneous emission may become asymmetric with respect to opposite directions. This directional effect is due to spin-orbit coupling of light carrying transverse spin angular momentum  \cite{Zeldovich,Bliokh review,Bliokh2014,Bliokh review2015,Bliokh2015,Banzer review2015,Lodahl2017}. Asymmetric spontaneous emission may lead to a nonzero average spontaneous emission recoil and, hence, may contribute to the optical force on the atoms. In particular, a lateral spontaneous emission recoil force may arise for an initially excited atom near a nanofiber \cite{Scheel2015,Jacob2016}, a flat surface \cite{Jacob2016,Buhmann2018}, or a photonic topological material \cite{Anzetta2018a,Anzetta2018b}. Such a lateral force appears because, in the presence of a material object, the interaction between the radiation field and the atom is chiral \cite{Fam2014,Petersen2014,Mitsch14b,sponhigh,Scheel2015,Jacob2016,flat,Buhmann2018,Anzetta2018a,Anzetta2018b,leFeberScience2015,Fuchs2017}. For an atom driven by a guided field,  
the spontaneous emission rate and the Rabi frequency may depend on the field propagation direction.  The effects of the directional dependencies of the spontaneous emission rate and the Rabi frequency on the optical force for an atom near an ultrathin optical fiber have recently been studied \cite{chiralforce}. The Casimir-Polder potential of an atom driven by a laser field near a flat surface has been calculated \cite{Buhmann2018a}.
It is worth noting that asymmetric  coupling  not only allows one to selectively excite modes in a preferential direction but also leads to effects like modified superradiance and subradiance \cite{KienPRA2017, KornovanPRB2017}, nonreciprocal transmission \cite{JungePRL2013}, and modified strong-coupling regime \cite{ChevryACSPH2018}.

The aim of this paper is to present a significant extension and comprehensive treatment for the force of light on a two-level atom near an ultrathin optical fiber. We calculate analytically and numerically all the components of the force of light. 
Furthermore, in this paper we use the mode function method as well as the Green function technique and show the connection between them. This gives us access to more details and broader insights. In particular, we compute the van der Waals potentials for the atom in the ground and excited states.

The paper is organized as follows. In Sec.~\ref{sec:model} we describe the model system.
Section \ref{sec:force} is devoted to deriving the expressions for the force in terms of the mode functions and the Green tensor.
In Sec.~\ref{sec:numerical} we present numerical results. 
Our conclusions are given in Sec.~\ref{sec:summary}.

\section{Model}
\label{sec:model}

\begin{figure}[tbh]
\begin{center}
  \includegraphics{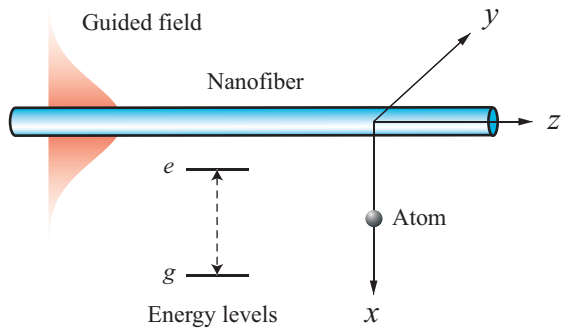}
 \end{center}
\caption{ A two-level atom driven by guided light of an ultrathin optical fiber. 
}
\label{fig1}
\end{figure}

We consider a two-level atom driven by a classical field in a guided mode of a vacuum-clad ultrathin optical fiber (see figure \ref{fig1}). 
The atom has an upper energy level $|e\rangle$ and a lower energy level $|g\rangle$, with energies $\hbar\omega_e$ and $\hbar\omega_g$, respectively. 
The atomic transition frequency is $\omega_0=\omega_e-\omega_g$. The fiber is a dielectric cylinder of radius $a$ and refractive index $n_1>1$ and is surrounded by an infinite background vacuum or air medium of refractive index $n_2=1$. We use Cartesian coordinates $\{x,y,z\}$, where $z$ is the coordinate along the fiber axis, and also cylindrical coordinates $\{r,\varphi,z\}$, where $r$ and $\varphi$ are the polar coordinates in the fiber transverse plane $xy$. 
In addition to the classical guided driving field, the quantum electromagnetic field 
interacts with the atom leading to spontaneous emission and energy level shift.

%%%%%%%%%%%%%%%%%%%%%%% Figure 1

\subsection{Quantum electromagnetic field}

The positive-frequency part $\mathbf{E}^{(+)}$ of the electric component of the field can be
decomposed into the contributions $\mathbf{E}^{(+)}_{\mathrm{g}}$ and $\mathbf{E}^{(+)}_{\mathrm{r}}$ from guided and radiation modes, respectively,  as  
\begin{equation}
\mathbf{E}^{(+)}=\mathbf{E}^{(+)}_{\mathrm{g}}+\mathbf{E}^{(+)}_{\mathrm{r}}.
\label{c1}
\end{equation}
In view of the very low losses of silica in the wavelength range of interest, we neglect material absorption. 

Regarding the guided modes, we assume that the fiber supports the fundamental HE$_{11}$ mode and a few higher-order modes \cite{fiber books} 
in a finite bandwidth around the atomic transition frequency $\omega_0$. 
We label each guided mode in this bandwidth by an index $\mu=(\omega N f p)$. 
Here, $\omega$ is the mode frequency, the notation $N=\mathrm{HE}_{lm}$, EH$_{lm}$, TE$_{0m}$, or TM$_{0m}$ stands for the mode type, with $l=1,2,\dots$
being the azimuthal order and $m=1,2,\dots$ being the radial mode order, the index $f=+1$ or $-1$ denotes the forward or backward propagation direction along the fiber axis $z$, and $p$ is the polarization index. The HE$_{lm}$ and EH$_{lm}$ modes are hybrid modes. For these modes, the azimuthal order is $l\not=0$,
and the index $p$ is equal to $+1$ or $-1$, indicating the counterclockwise or clockwise circulation direction of the helical phasefront. The TE$_{0m}$ and TM$_{0m}$ modes are transverse electric and magnetic modes. For these modes, the azimuthal mode order is $l=0$ and, hence, the mode polarization is single
and the polarization index $p$ can take an arbitrary value. For convenience, we assign the value $p=0$ to the polarization index $p$ for TE$_{0m}$ and TM$_{0m}$ modes. In the interaction picture, the quantum expression for the positive-frequency part $\mathbf{E}^{(+)}_{\mathrm{g}}$ of the electric component of the field in guided modes is \cite{sponhigh}
\begin{equation}
\mathbf{E}^{(+)}_{\mathrm{g}}=\rmi\sum_{\mu}\sqrt{\frac{\hbar\omega\beta'}{4\pi\epsilon_0}}
\;a_{\mu}\mathbf{e}^{(\mu)}\rme^{-\rmi(\omega t-f\beta z-pl\varphi)}.
\label{c2}
\end{equation}
Here, $\mathbf{e}^{(\mu)}=\mathbf{e}^{(\mu)}(r,\varphi)$ is the profile function of the guided mode $\mu$ in the classical problem, $a_{\mu}$ is the corresponding photon annihilation  operator, 
$\sum_{\mu}=\sum_{N fp}\int_0^{\infty}\rmd\omega$ is the generalized summation over the guided modes,
$\beta$ is the longitudinal propagation constant, and $\beta'$ is the derivative of $\beta$
with respect to $\omega$. The constant $\beta$ is determined by the
fiber eigenvalue equation \cite{fiber books}. The operators $a_{\mu}$ and $a_{\mu}^\dagger$ satisfy the continuous-mode bosonic commutation rules $[a_{\mu},a_{\mu'}^\dagger]=\delta(\omega-\omega')\delta_{NN'}\delta_{ff'}\delta_{pp'}$. 
The normalization condition for the guided mode profile function $\mathbf{e}^{(\mu)}$ is 
\begin{equation}\label{c3}
\int _{0}^{2\pi}\rmd\varphi\int _{0}^{\infty}n_{\mathrm{ref}}^2\,|\mathbf{e}^{(\mu)}|^2r\rmd r=1,
\end{equation}
where $n_{\mathrm{ref}}(r)=n_1$ for $r<a$ and $n_2$ for $r>a$.
The explicit expressions for the profile functions $\mathbf{e}^{(\mu)}$ of guided modes are given in  \cite{fiber books,highorder}. An important property of the mode functions of hybrid and TM modes is that the longitudinal
component $e_z$ is nonvanishing and in quadrature ($\pi/2$ out of phase) with the radial component $e_r$.

For radiation modes, the longitudinal propagation constant $\beta$ for each value of the frequency $\omega$ can vary continuously, from $-kn_2$ to $kn_2$ (with $k=\omega/c$). We label each radiation mode by an index $\nu=(\omega \beta l p)$, where 
$l=0,\pm1,\pm2,\dots$ is the mode order and  $p=+,-$ is the mode polarization. In the interaction picture, the quantum expression for the positive-frequency part $\mathbf{E}^{(+)}_{\mathrm{r}}$ of the electric component of the field in radiation modes is \cite{sponhigh}
\begin{equation}
\mathbf{E}^{(+)}_{\mathrm{r}}=\rmi\sum_{\nu}
\sqrt{\frac{\hbar\omega}{4\pi\epsilon_0}}\;a_{\nu}\mathbf{e}^{(\nu)}\rme^{-\rmi(\omega t-\beta z-l\varphi)}.
\label{c7}
\end{equation}
Here, $\mathbf{e}^{(\nu)}=\mathbf{e}^{(\nu)}(r,\varphi)$ is the profile function of the radiation mode $\nu$ in the classical problem, $a_{\nu}$ is the corresponding photon annihilation 
operator, and $\sum_{\nu}=\sum_{lp}\int_0^{\infty}\rmd\omega\int_{-kn_2}^{kn_2}\rmd\beta$ is the generalized summation over the radiation modes. The operators $a_{\nu}$ and $a_{\nu}^\dagger$ satisfy the continuous-mode bosonic commutation rules $[a_{\nu},a_{\nu'}^\dagger]=\delta(\omega-\omega')\delta(\beta-\beta')
\delta_{ll'}\delta_{pp'}$. 
The normalization condition for the radiation mode profile function $\mathbf{e}^{(\nu)}$ is 
\begin{equation}\label{c8}
\int _0^{2\pi}\rmd\varphi\int _{0}^{\infty}n_{\mathrm{ref}}^2
\left[\mathbf{e}^{(\nu)}\mathbf{e}^{(\nu')*}\right]_{\beta=\beta',l=l',p=p'}
r\rmd r=\delta(\omega-\omega').
\end{equation}
The explicit expressions for the mode functions $\mathbf{e}^{(\nu)}$ are given in \cite{sponhigh,fiber books}.

\subsection{Classical guided driving field}

We describe the classical guided driving field. We assume that the driving field is prepared in a hybrid HE or EH mode, a TE mode, or a TM mode.
Let $\omega_L$ be the frequency of the field.
For a quasicircularly hybrid $\mathrm{HE}_{lm}$ or EH$_{lm}$ mode with propagation direction $f_L$ and phase circulation direction $p_L$, the field amplitude is 
\begin{equation}\label{c9}
\boldsymbol{\mathcal{E}}=\mathcal{A} (e_r\hat{\mathbf{r}}+p_Le_\varphi\hat{\boldsymbol{\varphi}}+f_Le_z\hat{\mathbf{z}})\rme^{\rmi f_L\beta_L z+\rmi p_Ll\varphi},
\end{equation}
where $\mathcal{A}$ is a constant. For a TE$_{0m}$ mode with propagation direction $f_L$, the field amplitude is
\begin{equation}\label{c10}
\boldsymbol{\mathcal{E}}=\mathcal{A} e_\varphi\hat{\boldsymbol{\varphi}}\rme^{\rmi f_L\beta_L z}.
\end{equation}
For a TM mode with propagation direction $f_L$, the field amplitude is
\begin{equation}\label{c11}
\boldsymbol{\mathcal{E}}=\mathcal{A} (e_r\hat{\mathbf{r}}+f_Le_z\hat{\mathbf{z}})\rme^{\rmi f_L\beta_L z}.
\end{equation}
Quasilinearly polarized hybrid modes are linear superpositions of counterclockwise and clockwise quasicircularly polarized hybrid modes. The amplitude of the guided field in a quasilinearly polarized hybrid mode can be written in the form 
\begin{eqnarray}\label{c12}
\boldsymbol{\mathcal{E}}&=&\sqrt2\mathcal{A} [e_r\cos (l\varphi-\varphi_{\mathrm{pol}})\,\hat{\mathbf{r}}
+\rmi e_\varphi\sin (l\varphi-\varphi_{\mathrm{pol}})\,\hat{\boldsymbol{\varphi}}
+f_Le_z\cos (l\varphi-\varphi_{\mathrm{pol}})\,\hat{\mathbf{z}}]
\nonumber\\&&\mbox{}\times
\rme^{\rmi f_L\beta_L z},
\end{eqnarray}
where the phase angle $\varphi_{\mathrm{pol}}$ determines the orientation of the symmetry axes of the mode profile in the fiber transverse plane. In particular, the specific values $\varphi_{\mathrm{pol}}=0$ and $\pi/2$ define two orthogonal polarization profiles, called even and odd, respectively.
In equations  (\ref{c9})--(\ref{c12}), the mode profile function components $e_r$, $e_\varphi$, and $e_z$ are evaluated at $\omega=\omega_L$ and $\beta=\beta_L$.

\subsection{Atom--field interaction}

We introduce the atomic operators $\sigma_{ij}=|i\rangle\langle j|$, where $i,j=e,g$. 
The operators $\sigma_{eg}=|e\rangle\langle g|$ and $\sigma_{ge}=|g\rangle\langle e|$ describe the upward and downward transitions, respectively.
The operators $\sigma_{ee}=|e\rangle\langle e|$ and $\sigma_{gg}=|g\rangle\langle g|$ describe the populations of the upper and lower levels, respectively. 
We denote the position of the atom as $(r,\varphi,z)$. 

The Hamiltonian for the atom-field interaction in the dipole approximation is given by 
\begin{eqnarray}\label{c13}
H_{\mathrm{int}}&=&-\frac{\hbar}{2}\Omega\sigma_{eg}\rme^{-\rmi(\omega_L-\omega_0)t}
-\rmi\hbar\sum_{\alpha}G_{\alpha}\sigma_{eg} a_{\alpha}\rme^{-\rmi(\omega-\omega_0)t}
\nonumber\\&&\mbox{}
-\rmi\hbar\sum_{\alpha}\tilde{G}_{\alpha}\sigma_{ge} a_{\alpha}\rme^{-\rmi(\omega+\omega_0)t}
+\mbox{H.c.},
\end{eqnarray}
where $\Omega=\mathbf{d}\cdot\boldsymbol{\mathcal{E}}/\hbar$ is the Rabi frequency,
the notations $\alpha=\mu,\nu$ and $\sum_{\alpha}=\sum_{\mu}+\sum_{\nu}$   
stand for the general mode index and the full mode summation, respectively, and
the coefficients 
\begin{eqnarray}\label{c14}
G_{\mu}&=&\sqrt{\frac{\omega\beta'}{4\pi\epsilon_0\hbar}}\;
(\mathbf{d}\cdot\mathbf{e}^{(\mu)})\rme^{\rmi (f\beta z+pl\varphi)},\nonumber\\
G_{\nu}&=&\sqrt{\frac{\omega}{4\pi\epsilon_0\hbar}}\;
(\mathbf{d}\cdot\mathbf{e}^{(\nu)})\rme^{\rmi (\beta z+l\varphi)},
\end{eqnarray}
and
\begin{eqnarray}\label{c15}
\tilde{G}_{\mu}&=&\sqrt{\frac{\omega\beta'}{4\pi\hbar\epsilon_0}}\;
(\mathbf{d}^*\cdot\mathbf{e}^{(\mu)})\rme^{\rmi (f\beta z+pl\varphi)},\nonumber\\
\tilde{G}_{\nu}&=&\sqrt{\frac{\omega}{4\pi\hbar\epsilon_0}}\;
(\mathbf{d}^*\cdot\mathbf{e}^{(\nu)})\rme^{\rmi (\beta z+l\varphi)}
\end{eqnarray}
characterize the coupling of the atom with
the guided mode $\mu$ and the radiation mode $\nu$. Here, $\mathbf{d}=\langle e|\mathbf{D}|g\rangle$ is the matrix element of the atomic dipole operator $\mathbf{D}$. The coefficient $G_{\alpha}$ characterizes the coupling of the atom  with
mode $\alpha$ via the corotating term $\sigma_{eg} a_\alpha$.  
The coefficient $\tilde{G}_{\alpha}$ describes the coupling of the atom with
mode $\alpha$ via the counterrotating term $\sigma_{ge} a_\alpha$.
In deriving the Hamiltonian (\ref{c13}) we have used the rotating-wave approximation for the driving field but not for the quantum field. 
 
\section{Radiation force on an atom}
\label{sec:force}

The interaction between an atom and the light field affects the internal state
of the atom and leads to a radiation force.

\subsection{Excitation of an atom}

We consider the excitation of an atom. We call $\rho^{(I)}$
the density operator of the atomic internal state in the interaction picture.
We introduce the phase-shifted density operator $\rho$ with the matrix elements $\rho_{ee}=\rho_{ee}^{(I)}$, $\rho_{gg}=\rho_{gg}^{(I)}$,
$\rho_{ge}=\rho_{ge}^{(I)}\rme^{-\rmi(\omega_L-\omega_0)t}$, and $\rho_{eg}=\rho_{eg}^{(I)}\rme^{\rmi (\omega_L-\omega_0)t}$. 
We obtain the generalized Bloch equations \cite{coolingbook} 
\begin{eqnarray}\label{c16}
\dot{\rho}_{ee}&=&\frac{\rmi}{2}(\Omega\rho_{ge}
-\Omega^*\rho_{eg})-\Gamma{\rho}_{ee},\nonumber\\
\dot{\rho}_{gg}&=&-\frac{\rmi}{2}(\Omega
\rho_{ge}-\Omega^*\rho_{eg})+\Gamma{\rho}_{ee},\nonumber\\
\dot{\rho}_{ge}&=&\frac{\rmi}{2}\Omega^*(\rho_{ee}-\rho_{gg})-\bigg(\frac{\Gamma}{2}+\rmi\Delta\bigg)\rho_{ge}.
\end{eqnarray}
Here, $\Delta=\omega_L-\tilde{\omega}_0$ is the detuning of the frequency $\omega_L$ of the driving field 
from the frequency $\tilde{\omega}_0=\omega_0+\delta\omega_0$ of the atomic transition between the shifted levels, with \cite{Eberly} 
\begin{equation}\label{c17}
\delta\omega_0=-\mathcal{P}\sum_{\alpha}\left(\frac{|G_{\alpha}|^2}{\omega-\omega_0}-\frac{|G_{\alpha}|^2}{\omega+\omega_0}\right).
\end{equation}
The parameter $\Gamma=\gamma_{\mathrm{g}}+\gamma_{\mathrm{r}}$ is the rate of spontaneous emission, with \cite{sponhigh}
\begin{equation}\label{c18}
\gamma_{\mathrm{g}}=2\pi \sum_{Nfp}|G_{\omega_0Nfp}|^2
\end{equation}
and
\begin{equation}\label{c19}
\gamma_{\mathrm{r}}=2\pi \sum_{lp}\int_{-k_0n_2}^{k_0n_2}|G_{\omega_0\beta lp}|^2 \rmd\beta
\end{equation}
being the contributions from the resonant guided and radiation modes, respectively.

We consider the regime where the atom is at rest and in the steady state.
In this regime, we can set the derivatives in equations  (\ref{c16}) to zero. Then, we obtain  \cite{coolingbook}
\begin{eqnarray}\label{c40}
\rho_{ee}&=&\frac{1}{2}\frac{s}{1+s}=\frac{|\Omega|^2/4}{\Delta^2+\Gamma^2/4+|\Omega|^2/2},\nonumber\\
\rho_{eg}&=&\frac{\rmi\Omega}{(\Gamma-2\rmi\Delta)(1+s)}=\frac{\rmi\Omega(\Gamma+2\rmi\Delta)/4}{\Delta^2+\Gamma^2/4+|\Omega|^2/2},
\end{eqnarray}
where
\begin{equation}\label{c41}
s=\frac{|\Omega|^2/2}{\Delta^2+\Gamma^2/4}
\end{equation}
is the saturation parameter.

\subsection{Force on an atom in terms of the mode functions}

We consider the center-of-mass motion of the atom and perform a semiclassical treatment
for this motion.
In such a treatment, the center-of-mass motion is governed by the force calculated
from the quantum internal state of the atom.
The force of the light field on the atom is defined by the formula
\begin{equation}\label{c20}
\mathbf{F}= -\langle\boldsymbol{\nabla} H_{\mathrm{int}}\rangle.
\end{equation}
We use the interaction picture.
Inserting equation (\ref{c13}) into equation (\ref{c20}) gives the following expression for the force:
\begin{eqnarray}\label{c21}
\mathbf{F}&=&\bigg\{\frac{\hbar}{2}(\boldsymbol{\nabla}\Omega)\langle \sigma_{eg} \rangle \rme^{-\rmi(\omega_L-\omega_0)t}
+\rmi\hbar\sum_{\alpha}(\boldsymbol{\nabla}G_{\alpha})\langle \sigma_{eg} a_{\alpha} \rangle \rme^{-\rmi(\omega-\omega_0)t}
\nonumber\\&&\mbox{}
+\rmi\hbar\sum_{\alpha}(\boldsymbol{\nabla}\tilde{G}_{\alpha})\langle \sigma_{ge} a_{\alpha} \rangle \rme^{-\rmi(\omega+\omega_0)t}+\mathrm{c.c.}\bigg\}.
\end{eqnarray}
Meanwhile, the Heisenberg equation for the photon operator $a_{\alpha}$ is
$\dot{a}_{\alpha}=G_{\alpha}^*\sigma_{ge}\rme^{\rmi (\omega-\omega_0)t}+\tilde{G}_{\alpha}^*\sigma_{eg} \rme^{\rmi (\omega+\omega_0)t}$.
Integrating  this equation, we find
\begin{equation}\label{c22}
a_{\alpha}(t)=a_{\alpha}(t_0)+G_{\alpha}^*\int_{t_0}^t \rmd t'\,\sigma_{ge}(t')\rme^{\rmi (\omega-\omega_0)t'}
+\tilde{G}_{\alpha}^*\int\limits _{t_0}^t \rmd t'\,\sigma_{eg}(t')\rme^{\rmi (\omega+\omega_0)t'},
\end{equation}
where $t_0$ is the initial time. In deriving equation (\ref{c22}), we have neglected the time dependence of the position of the atom. 
We consider the situation where the quantum electromagnetic field is initially in the vacuum state. 
We assume that the evolution time $t-t_0$ and the characteristic atomic lifetime $\tau$ are 
large as compared to the characteristic optical period $T=2\pi/\omega_0$. 
Under these conditions, since the continuum of the field modes is broadband and the interaction between the atom and the field is weak,
the Born-Markov approximation $\sigma_{ge}(t')=\sigma_{ge}(t)$ can be applied to describe the back
action of the second and third terms in equation (\ref{c22}) on the atom \cite{Eberly}. 
Under the condition $t-t_0\gg T$, we calculate the integral with respect to $t'$ in the limit $t-t_0\to\infty$.  
With the above approximations, we obtain
\begin{eqnarray}\label{c23}
a_{\alpha}(t)&=&a_{\alpha}(t_0)+\pi G_{\alpha}^*\sigma_{ge}(t)\delta(\omega-\omega_0)
%\nonumber\\&&\mbox{}
-\rmi G_{\alpha}^*\sigma_{ge}(t)\rme^{\rmi (\omega-\omega_0)t}\frac{\mathcal{P}}{\omega-\omega_0}
\nonumber\\&&\mbox{}
-\rmi\tilde{G}_{\alpha}^*\sigma_{eg}(t)\rme^{\rmi (\omega+\omega_0)t}\frac{\mathcal{P}}{\omega+\omega_0},
\end{eqnarray}
where the notation $\mathcal{P}$ stands for the principal value.
We substitute equation (\ref{c23}) into equation (\ref{c21}) and neglect fast-oscillating terms. 
With the use of the relations $\rho_{ee}=\langle\sigma_{ee}\rangle$, $\rho_{gg}=\langle\sigma_{gg}\rangle$,
$\rho_{ge}=\langle\sigma_{eg}\rangle \rme^{-\rmi(\omega_L-\omega_0)t}$, and $\rho_{eg}=\langle\sigma_{ge}\rangle \rme^{\rmi (\omega_L-\omega_0)t}$,
we obtain  \cite{chiralforce}
\begin{equation}\label{c25}
\mathbf{F}=\mathbf{F}^{\mathrm{(drv)}}+\rho_{ee}\mathbf{F}^{\mathrm{(spon)}}+\rho_{ee}\mathbf{F}^{(\mathrm{vdW})e}
+\rho_{gg}\mathbf{F}^{(\mathrm{vdW})g},             
\end{equation}
where
\begin{equation}\label{c26}
\mathbf{F}^{\mathrm{(drv)}}=\frac{\hbar}{2}(\rho_{ge}\boldsymbol{\nabla}\Omega+\rho_{eg}\boldsymbol{\nabla}\Omega^*)             
\end{equation}
is the force resulting from the interaction with the driving field,
\begin{equation}\label{c27}
\mathbf{F}^{\mathrm{(spon)}}=\rmi\pi\hbar\sum_{\alpha_0}(G_{\alpha_0}^*\boldsymbol{\nabla}G_{\alpha_0}-G_{\alpha_0}\boldsymbol{\nabla}G_{\alpha_0}^*)    \end{equation}
is the force resulting from the recoil of spontaneous emission of the atom in the excited state \cite{Scheel2015}, and
\begin{equation}\label{c28}
\mathbf{F}^{(\mathrm{vdW})e}=\hbar\boldsymbol{\nabla}\mathcal{P}\sum_{\alpha}\frac{|G_{\alpha}|^2}{\omega-\omega_0}
\end{equation}
and
\begin{equation}\label{c29}
\mathbf{F}^{(\mathrm{vdW})g}=\hbar\boldsymbol{\nabla}\mathcal{P}\sum_{\alpha}\frac{|G_{\alpha}|^2}{\omega+\omega_0}    \end{equation}
are the forces resulting from the van der Waals potentials for the excited and ground states, respectively.
In equation (\ref{c27}), the notation $\alpha_0$ is the label of a resonant guided mode $\mu_0=(\omega_0 N f p)$ or a resonant radiation mode $\nu_0=(\omega_0 \beta l p)$, and the generalized summation $\sum_{\alpha_0}$ is defined as $\sum_{\alpha_0}=\sum_{\mu_0}+\sum_{\nu_0}$ with
$\sum_{\mu_0}=\sum_{N fp}$ and $\sum_{\nu_0}=\sum_{lp}\int_{-k_0n_2}^{k_0n_2}\rmd\beta$. We note that $\mathbf{F}^{\mathrm{(spon)}}$ and $\mathbf{F}^{(\mathrm{vdW})e}$ enter equation (\ref{c25}) with the weight factor $\rho_{ee}$, while $\mathbf{F}^{(\mathrm{vdW})g}$ enters with the weight factor $\rho_{gg}$.
The term $\mathbf{F}^{(\mathrm{scatt})}\equiv\rho_{ee}\mathbf{F}^{\mathrm{(spon)}}$ is the force produced by the recoil of the photons that are scattered from the atom with the excited-state population $\rho_{ee}$.
In deriving equation (\ref{c29}) we have used the symmetry property $|\tilde{G}_{\alpha}|^2=|G_{\tilde{\alpha}}|^2$,
where $\tilde{\alpha}=\tilde{\mu}=(\omega,N,-f,-p)$ for $\alpha=\mu=(\omega,N,f,p)$ in the case of guided modes
and $\tilde{\alpha}=\tilde{\nu}=(\omega,-\beta,-l,p)$ for $\alpha=\nu=(\omega,\beta,l,p)$ in the case of radiation modes \cite{sponhigh}.

The force $\mathbf{F}^{\mathrm{(drv)}}$ of the driving field includes the effects of the momentum transfers in the competing elementary absorption and stimulated emission processes. This force also includes the effect of the AC-Stark shifts of the atomic energy levels.

The forces $\mathbf{F}^{(\mathrm{vdW})e}$ and $\mathbf{F}^{(\mathrm{vdW})g}$ are produced by the van der Waals potentials $U_e$ and $U_g$ \cite{Buhmann2004}, that is, $\mathbf{F}^{(\mathrm{vdW})e}=-\boldsymbol{\nabla} U_e$ and $\mathbf{F}^{(\mathrm{vdW})g}=-\boldsymbol{\nabla} U_g$. 
These body-induced potentials are given as
\begin{eqnarray}\label{c30}
U_e&=&-\hbar\mathcal{P}\sum_{\alpha}\frac{|G_{\alpha}|^2}{\omega-\omega_0}-\delta E_e^{(\mathrm{vac})},
\nonumber\\
U_g&=&-\hbar\mathcal{P}\sum_{\alpha}\frac{|G_{\alpha}|^2}{\omega+\omega_0}-\delta E_g^{(\mathrm{vac})},  
\end{eqnarray}
where $\delta E_e^{(\mathrm{vac})}$ and $\delta E_g^{(\mathrm{vac})}$ are the energy level shifts induced by the vacuum field in free space (in the absence of the fiber).
Note that $\delta E_e^{(\mathrm{vac})}-\delta E_g^{(\mathrm{vac})}=\hbar\delta\omega_0^{(\mathrm{vac})}$, where $\delta\omega_0^{(\mathrm{vac})}$ is the Lamb shift of the transition frequency of the atom in free space. 
The detuning of the field from the atom near the fiber can be written as $\Delta=\Delta_0-(U_e-U_g)/\hbar$, where $\Delta_0=\omega_L-\omega_0-\delta\omega_0^{(\mathrm{vac})}$ is the detuning of the field from the atom in the absence of the fiber.

We now calculate the individual components of the force.
When we use the symmetry of the mode profile functions, we find $F_z^{(\mathrm{vdW})e}=F_z^{(\mathrm{vdW})g}=0$.
Then, the axial component $F_z$ of the total force is found to be 
\begin{equation}\label{c35}
F_z=F_z^{\mathrm{(drv)}}+\rho_{ee}F_z^{\mathrm{(spon)}},
\end{equation}
where
\begin{eqnarray}\label{c35a}
F_z^{\mathrm{(drv)}}&=&\frac{\rmi\hbar f_L\beta_L}{2}(\Omega\rho_{ge}-\Omega^*\rho_{eg}),
\nonumber\\
F_z^{\mathrm{(spon)}}&=&-\sum_{N}\hbar\beta_{0}^{(N)}(\gamma_{\mathrm{g}N}^{(+)}-\gamma_{\mathrm{g}N}^{(-)})
-\int_{-k_0n_2}^{k_0n_2}\hbar\beta\gamma^{(\beta)}_{\mathrm{r}} \rmd\beta.
\end{eqnarray}
Here, we have introduced the notation
\begin{equation}\label{c36}
\gamma_{\mathrm{g}N}^{(f)}=2\pi \sum_{p}|G_{\omega_0Nfp}|^2
\end{equation}
for the rate of spontaneous emission into the guided modes of type $N$ with the propagation direction $f=\pm$, 
and the notation
\begin{equation}\label{c37}
\gamma^{(\beta)}_{\mathrm{r}}=2\pi\sum_{lp}|G_{\omega_0\beta lp}|^2
\end{equation}
for the rate of spontaneous emission into the radiation modes with the axial component $\beta$ of the wave vector. 
It is clear that $F_z^{\mathrm{(drv)}}$ is the recoil of the absorbed photons 
and $\rho_{ee}F_z^{\mathrm{(spon)}}$ is the recoil of the photons re-emitted into guided and radiation modes.
The component $F_z^{\mathrm{(drv)}}$ is a light pressure force \cite{coolingbook}.

We can show that $F_z^{\mathrm{(spon)}}\propto \mathrm{Im}\,[d_r^*d_z]$. 
Thus, $F_z^{\mathrm{(spon)}}$ is nonzero when $\mathrm{Im}\,[d_r^*d_z]\not=0$, that is, when the atomic dipole rotates in the meridional plane $rz$.
In the case where $\mathrm{Im}\,[d_r^*d_z]=0$, spontaneous emission is symmetric with respect to the forward and backward directions. 
In this case, we have $F_z^{\mathrm{(spon)}}=0$ and hence $F_z=F_z^{\mathrm{(drv)}}$.

For the atom in the steady-state regime, we find the expression $F_z^{\mathrm{(drv)}}=f_L\hbar \beta_L\Gamma\rho_{ee}$. In this case, we have
\begin{equation}\label{c37a}
F_z=\hbar\rho_{ee}\bigg\{f_L\beta_L\Gamma
-\sum_{N}\beta_0^{(N)}(\gamma_{\mathrm{g}N}^{(+)}-\gamma_{\mathrm{g}N}^{(-)})
-\int\limits_{-k_0n_2}^{k_0n_2}\beta\gamma^{(\beta)}_{\mathrm{r}} \rmd\beta\bigg\}.
\end{equation}

Making use of the symmetry properties of the mode functions, we can show that $F_r^{\mathrm{(spon)}}=0$.
Then, the radial component $F_r$ of the total force is found to be
\begin{equation}\label{c38}
F_r=F_r^{\mathrm{(drv)}}+\rho_{ee}F_r^{(\mathrm{vdW})e}+\rho_{gg}F_r^{(\mathrm{vdW})g},
\end{equation}
where
\begin{eqnarray}\label{c38a}
F_r^{\mathrm{(drv)}}&=&\frac{\hbar}{2}\left(\rho_{ge}\frac{\partial\Omega}{\partial r}+\rho_{eg}\frac{\partial\Omega^*}{\partial r}\right),
\nonumber\\
F_r^{(\mathrm{vdW})e}&=&-\frac{\partial U_e}{\partial r},\qquad
F_r^{(\mathrm{vdW})g}=-\frac{\partial U_g}{\partial r}.
\end{eqnarray}
Due to the evanescent-wave behavior of guided modes in the transverse plane, the radial component $F_r^{\mathrm{(drv)}}$ of the force of the driving field in a guided mode is a gradient force (dipole force) \cite{coolingbook}. 

Finally, we calculate the azimuthal component $F_\varphi$ of the total force. The result is  
\begin{equation}\label{c39}
F_\varphi=F_\varphi^{\mathrm{(drv)}}+\rho_{ee}F_\varphi^{\mathrm{(spon)}}
+\rho_{ee}F_\varphi^{(\mathrm{vdW})e}+\rho_{gg}F_\varphi^{(\mathrm{vdW})g},
\end{equation}
where
\begin{eqnarray}\label{c39a}
F_\varphi^{\mathrm{(drv)}}&=&\frac{\hbar}{2r}\left(\rho_{ge}\frac{\partial\Omega}{\partial \varphi}
+\rho_{eg}\frac{\partial\Omega^*}{\partial \varphi}\right),
\nonumber\\
F_\varphi^{\mathrm{(spon)}}&=&
\rmi\frac{\pi\hbar}{r}\sum_{\alpha_0}\left(G_{\alpha_0}^*\frac{\partial G_{\alpha_0}}{\partial \varphi}
-G_{\alpha_0}\frac{\partial G_{\alpha_0}^*}{\partial \varphi}\right),
\nonumber\\
F_\varphi^{(\mathrm{vdW})e}&=&-\frac{1}{r}\frac{\partial U_e}{\partial \varphi},\qquad
F_\varphi^{(\mathrm{vdW})g}=-\frac{1}{r}\frac{\partial U_g}{\partial \varphi}.
\end{eqnarray}
It is clear that the azimuthal component $F_\varphi^{\mathrm{(drv)}}$ of the driving-field force is determined by the gradient of the Rabi frequency of the driving field with respect to the azimuthal angle $\varphi$. This component is, in general, a combination of the pressure and gradient forces in the azimuthal direction \cite{coolingbook}.
We can show that $F_\varphi^{\mathrm{(spon)}}\propto \mathrm{Im}\,[d_r^*d_\varphi]$.
This result means that $F_\varphi^{\mathrm{(spon)}}$ is nonzero when $\mathrm{Im}\,[d_r^*d_\varphi]\not=0$, that is, when the atomic dipole rotates in the fiber transverse plane $xy$.

\subsection{Force in terms of the Green tensor}

Expressions (\ref{c27})--(\ref{c29}) describe the spontaneous-emission recoil force $\mathbf{F}^{\mathrm{(spon)}}$
and the van der Waals forces $\mathbf{F}^{(\mathrm{vdW})e}$ and $\mathbf{F}^{(\mathrm{vdW})g}$ in terms of the mode functions.
These forces can also be presented in terms of the Green tensor \cite{Buhmann2004,Dogariu2016}. The explicit expression for the Green tensor $\mathbf{G}$ of a two-layer fiber is given in  \cite{Li,Li03,Raabe}. The connection between the Green tensor and the mode functions is given in  \ref{sec:connection}.

With the help of equations (\ref{green4}) and (\ref{c14}), we can rewrite equation (\ref{c27}) for the spontaneous-emission recoil force $\mathbf{F}^{\mathrm{(spon)}}$ as
\begin{equation}\label{c33}
\mathbf{F}^{\mathrm{(spon)}}=
\frac{\rmi\omega_0^2}{\epsilon_0 c^2}\boldsymbol{\nabla}\{\mathbf{d}\cdot\mathrm{Im}[\mathbf{G}^{(\mathrm{R})}(\mathbf{R},\mathbf{R}';\omega_0)]\cdot\mathbf{d}^*\}\big|_{\mathbf{R}'=\mathbf{R}}+\mathrm{c.c.},
\end{equation}
where $\mathbf{G}^{(\mathrm{R})}$ is the reflected part of the Green tensor. The equivalence of equations (\ref{c33}) and (\ref{c27}) can be easily verified by substituting equations  (\ref{green4}) into equation (\ref{c33}) and making use of equations (\ref{c14}).
It is clear from equation (\ref{c33}) that, when $\mathbf{d}$ is a real vector, that is, when the dipole of the atom is linearly polarized, we have $\mathbf{F}^{\mathrm{(spon)}}=0$.
However, when $\mathbf{d}$ is a complex vector, that is, when the dipole of the atom is elliptically polarized, we may obtain $\mathbf{F}^{\mathrm{(spon)}}\not=0$.

Similarly, with the help of equations (\ref{green4}) and (\ref{c14}), we can rewrite equations  (\ref{c30}) for the van der Waals potentials $U_e$ and $U_g$ as
\begin{eqnarray}\label{c31}
U_e&=&-\frac{1}{\pi\epsilon_0 c^2}\mathcal{P}
\int_0^{\infty}\rmd\omega\frac{\omega^2}{\omega-\omega_0}
\mathbf{d}\cdot\mathrm{Im}[\mathbf{G}^{(\mathrm{R})}(\mathbf{R},\mathbf{R};\omega)]\cdot\mathbf{d}^*,
\nonumber\\
U_g&=&-\frac{1}{\pi\epsilon_0 c^2}\mathcal{P}
\int_0^{\infty}\rmd\omega\frac{\omega^2}{\omega+\omega_0}
\mathbf{d}\cdot\mathrm{Im}[\mathbf{G}^{(\mathrm{R})}(\mathbf{R},\mathbf{R};\omega)]\cdot\mathbf{d}^*.
\end{eqnarray}
We can easily verify the equivalence of equations (\ref{c31}) and (\ref{c30}) by substituting equations  (\ref{green4}) into equation (\ref{c31}) and making use of equations (\ref{c14}).
It follows from the reciprocity property $G_{ij}^{(\mathrm{R})}(\mathbf{R},\mathbf{R}';\omega)=G_{ji}^{(\mathrm{R})}(\mathbf{R}',\mathbf{R};\omega)$
that $U_e$ and $U_g$ are real functions.

We use the contour integral technique to change the integrals in equations  (\ref{c31}) to the imaginary frequency.
Then, we obtain \cite{Buhmann2004}
\begin{eqnarray}\label{c32}
U_e&=&-\frac{\omega_0}{\pi\epsilon_0 c^2}
\int\limits_0^{\infty}\rmd u\frac{u^2}{\omega_0^2+u^2}
\mathbf{d}\cdot\mathrm{Re}[\mathbf{G}^{(\mathrm{R})}(\mathbf{R},\mathbf{R};\rmi u)]\cdot\mathbf{d}^*\nonumber\\
&&\mbox{}-\frac{\omega_0^2}{\epsilon_0 c^2}\mathbf{d}\cdot\mathrm{Re}[\mathbf{G}^{(\mathrm{R})}(\mathbf{R},\mathbf{R};\omega_0)]\cdot\mathbf{d}^*,
\nonumber\\
U_g&=&\frac{\omega_0}{\pi\epsilon_0 c^2}
\int\limits_0^{\infty}\rmd u\frac{u^2}{\omega_0^2+u^2}
\mathbf{d}\cdot\mathrm{Re}[\mathbf{G}^{(\mathrm{R})}(\mathbf{R},\mathbf{R};\rmi u)]\cdot\mathbf{d}^*.\nonumber\\
\end{eqnarray}
The first and second terms in the expression for $U_e$ in equations  (\ref{c32}) are respectively the off-resonant part $U_e^{(\mathrm{off})}$ and the resonant part $U_e^{(\mathrm{res})}$ of the van der Waals potential for the excited state $|e\rangle$ \cite{Buhmann2004}.
Thus, we can write $U_e=U_e^{(\mathrm{off})}+U_e^{(\mathrm{res})}$, where
\begin{eqnarray}\label{c32a}
U_e^{(\mathrm{off})}&=&-\frac{\omega_0}{\pi\epsilon_0 c^2}
\int\limits_0^{\infty}\rmd u\frac{u^2}{\omega_0^2+u^2}
\mathbf{d}\cdot\mathrm{Re}[\mathbf{G}^{(\mathrm{R})}(\mathbf{R},\mathbf{R};\rmi u)]\cdot\mathbf{d}^*,\nonumber\\
U_e^{(\mathrm{res})}&=&-\frac{\omega_0^2}{\epsilon_0 c^2}\mathbf{d}\cdot\mathrm{Re}[\mathbf{G}^{(\mathrm{R})}(\mathbf{R},\mathbf{R};\omega_0)]\cdot\mathbf{d}^*.
\end{eqnarray}
The potential $U_g$ for the ground state $|g\rangle$ does not contain a resonant part.
Note that $U_g$ is opposite to the off-resonant part $U_e^{(\mathrm{off})}$ of $U_e$, that is, $U_g=-U_e^{(\mathrm{off})}$.

Thus, expression (\ref{c25}) for the total radiation force $\mathbf{F}$ can be rewritten as
\begin{eqnarray}\label{c34}
\mathbf{F}&=&\mathbf{F}^{\mathrm{(drv)}}
+(\rho_{ee}-\rho_{gg})\frac{\omega_0}{\pi\epsilon_0 c^2}\int\limits_0^{\infty}\rmd u\frac{u^2}{\omega_0^2+u^2}
\boldsymbol{\nabla}\{\mathbf{d}\cdot\mathrm{Re}[\mathbf{G}^{(\mathrm{R})}(\mathbf{R},\mathbf{R};\rmi u)]\cdot\mathbf{d}^*\}\nonumber\\&&\mbox{}
+\rho_{ee}\frac{\omega_0^2}{\epsilon_0 c^2}
\big\{\boldsymbol{\nabla}[\mathbf{d}\cdot\mathbf{G}^{(\mathrm{R})}(\mathbf{R},\mathbf{R}';\omega_0)\cdot\mathbf{d}^*] 
\big|_{\mathbf{R}'=\mathbf{R}}
+\mathrm{c.c.}\big\}.\qquad         
\end{eqnarray}
Note that, since $\mathrm{Im}[\mathbf{G}^{(\mathrm{R})}(\mathbf{R},\mathbf{R};\rmi u)]=0$, we can replace $\mathrm{Re}[\mathbf{G}^{(\mathrm{R})}(\mathbf{R},\mathbf{R};\rmi u)]$ in equations  (\ref{c32})--(\ref{c34}) by $\mathbf{G}^{(\mathrm{R})}(\mathbf{R},\mathbf{R};\rmi u)$. 
The second term in equation (\ref{c34}) contains an integral over the imaginary frequency. This term describes the effects of the off-resonant van der Waals potentials $U_e^{(\mathrm{off})}$ and $U_g=-U_e^{(\mathrm{off})}$ on the force. The last term in equation (\ref{c34}) corresponds to 
the resonant excited-state van der Waals potential $U_e^{(\mathrm{res})}$ and the scattered-photon recoil.

Equation (\ref{c34}) is in agreement with the results of  \cite{Buhmann2004}, where multilevel atoms were considered. 
When we neglect the second term in equation (\ref{c34}), which corresponds to the off-resonant part of the van der Waals force, and assume the weak excitation regime, we can reduce equation (\ref{c34}) to 
\begin{eqnarray}\label{c45}
\mathbf{F}=\frac{1}{2}\sum_{i=x,y,z}\mathrm{Re}(\wp_i^*\boldsymbol{\nabla} \mathcal{E}_i)
+\frac{\omega_0^2\mu_0}{2}\sum_{i,j=x,y,z}\mathrm{Re}\{\wp_i^*\boldsymbol{\nabla}[ G_{ij}^{(\mathrm{R})}(\mathbf{R},\mathbf{R}';\omega_0)]\wp_j\}\big|_{\mathbf{R}'=\mathbf{R}},\nonumber\\
\end{eqnarray}
where $\boldsymbol{\wp}=\boldsymbol{\alpha}\boldsymbol{\mathcal{E}}$ is the positive frequency component of the induced dipole, with
\begin{equation}\label{c49}
\boldsymbol{\alpha}=-\frac{\mathbf{d}^*\mathbf{d}}{\hbar(\Delta+\rmi\Gamma/2)}
\end{equation}
being the fiber-enhanced atomic polarizability tensor.
Under the condition $\omega_L\simeq\omega_0$, equation (\ref{c45}) is in agreement with the results of  \cite{Dogariu2016} for classical point dipoles.

\section{Numerical calculations}
\label{sec:numerical}

We calculate numerically the force acting on the atom in the case where it is at rest and in the steady state.
We use the wavelength $\lambda_0=780$ nm and
the natural linewidth $\gamma_0/2\pi=6.065$ MHz, which correspond to the transitions in the $D_2$ line of  a $^{87}$Rb atom. 
The atomic dipole matrix element $d$ is calculated from the formula $\gamma_0=d^2\omega_0^3/3\pi\epsilon_0\hbar c^3$ for the natural linewidth of a two-level atom.
We assume that the fiber radius is $a=350$ nm, and the refractive indices of the fiber and the vacuum cladding are $n_1=1.4537$ and $n_2=1$, respectively.
The fiber can support the HE$_{11}$, TE$_{01}$,  TM$_{01}$, and HE$_{21}$ modes.
The atom is positioned on the $x$ axis if not otherwise specified. 
The driving field is prepared in a quasilinearly polarized hybrid HE mode, a TE mode, or a TM mode.
In the case of HE modes, we choose the $x$ polarization, which leads to a maximal longitudinal component of the field at the position of the atom.

\subsection{Driving-field force}

\begin{figure}[tbh]
\begin{center}
 \includegraphics{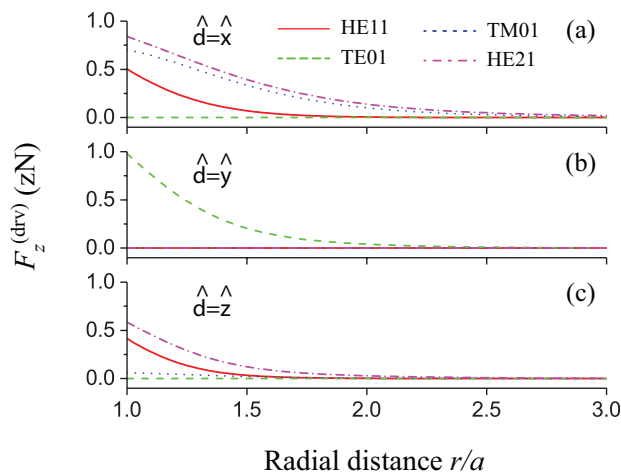}
 \end{center}
\caption{ Radial dependence of the axial component $F_z^{(\mathrm{drv})}$ of the driving-field force  $\mathbf{F}^{(\mathrm{drv})}$
 of a resonant ($\Delta=0$) guided light field on a two-level atom. 
The incident light field is in an $x$-polarized HE$_{11}$ mode (solid red curves), a TE$_{01}$ mode (dashed green curves), a TM$_{01}$ mode (dotted blue curves), or an $x$-polarized HE$_{21}$ mode (dashed-dotted magenta curves) and propagates in the forward $f_L=+1$ direction along the fiber axis $z$ with a power $P=1$ pW. The atom is positioned on the $x$ axis. The dipole orientation vector is $\hat{\mathbf{d}}=\hat{\mathbf{x}}$ (a), $\hat{\mathbf{y}}$ (b), and $\hat{\mathbf{z}}$ (c). The dipole magnitude $d$ corresponds to the natural linewidth $\gamma_0/2\pi=6.065$ MHz
of the $D_2$ line of a $^{87}$Rb atom. The fiber radius is $a=350$ nm. The wavelength of the atomic transition is $\lambda_0=780$ nm. The refractive indices of the fiber and the vacuum cladding are $n_1=1.4537$ and $n_2=1$, respectively.
The fiber-induced shift of the atomic transition frequency is neglected.}
\label{fig2}
\end{figure}

We first calculate the driving-field force $\mathbf{F}^{(\mathrm{drv})}$. 
We plot in figure \ref{fig2} the radial dependence of the axial component $F_z^{(\mathrm{drv})}$ of the driving-field force in the cases where
the driving field is at exact resonance with the atom ($\Delta=0$) and the dipole orientation vector $\hat{\mathbf{d}}\equiv\mathbf{d}/d$
coincides with one of the unit basis vectors $\hat{\mathbf{x}}$, $\hat{\mathbf{y}}$, and $\hat{\mathbf{z}}$ of the Cartesian coordinate system. 
As already mentioned in the previous section, $F_z^{(\mathrm{drv})}$ is  a pressure force.
Figure \ref{fig2} shows that $F_z^{(\mathrm{drv})}$ depends on the mode type and the orientation of the dipole vector.
We note that for the parameters of figure \ref{fig2}, the radial component $F_r^{(\mathrm{drv})}$ and the azimuthal component $F_\varphi^{(\mathrm{drv})}$ vanish and are therefore not plotted.
These transverse components of the driving-field force $\mathbf{F}^{(\mathrm{drv})}$ may appear when the dipole orientation vector is arbitrary
or the field detuning is nonzero.

%%%%%%%%%%%%%%%%%%%%%%% Figure 2

\begin{figure}[tbh]
\begin{center}
 \includegraphics{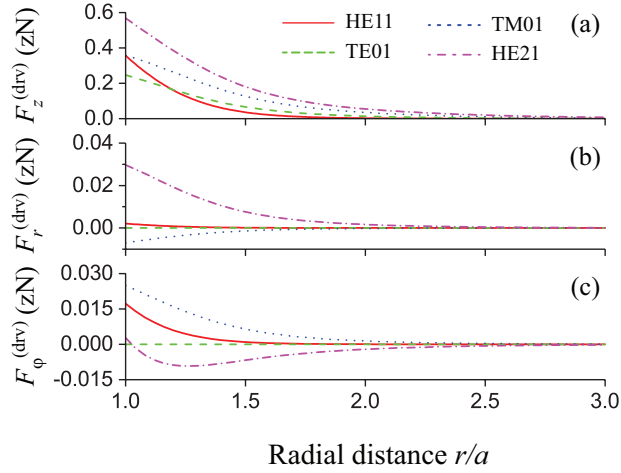}
 \end{center}
\caption{ Radial dependencies of the components $F_z^{(\mathrm{drv})}$ (a), $F_r^{(\mathrm{drv})}$ (b), and $F_\varphi^{(\mathrm{drv})}$ (c) of the driving-field force $\mathbf{F}^{(\mathrm{drv})}$
 of a resonant ($\Delta=0$) guided light field on the atom. 
The dipole orientation vector is $\hat{\mathbf{d}}=(\hat{\mathbf{x}}+\hat{\mathbf{y}}+\hat{\mathbf{z}})/\sqrt3$.
Other parameters are as for figure \ref{fig2}.}
\label{fig3}
\end{figure}

We show in figure \ref{fig3} the radial dependencies of the components 
$F_z^{(\mathrm{drv})}$, $F_r^{(\mathrm{drv})}$, and $F_\varphi^{(\mathrm{drv})}$
of the driving-field force in the case where the driving field is at exact resonance with the atom ($\Delta=0$) and the dipole orientation vector is $\hat{\mathbf{d}}=(\hat{\mathbf{x}}+\hat{\mathbf{y}}+\hat{\mathbf{z}})/\sqrt3$. 
We observe from figure \ref{fig3}(b) that, when the dipole is not strictly oriented along the $x$,
$y$, or $z$ direction, the radial force component $F_r^{(\mathrm{drv})}$ can be nonzero even though $\Delta=0$.
This feature is a consequence of the vector nature of the guided driving field. 
The dashed-dotted magenta curve in figure \ref{fig3}(c) indicates that the azimuthal force component $F_\varphi^{(\mathrm{drv})}$ for the HE$_{21}$ mode can be negative or positive depending on the radial position $r$.

%%%%%%%%%%%%%%%%%%%%%%% Figure 3

\begin{figure}[tbh]
\begin{center}
 \includegraphics{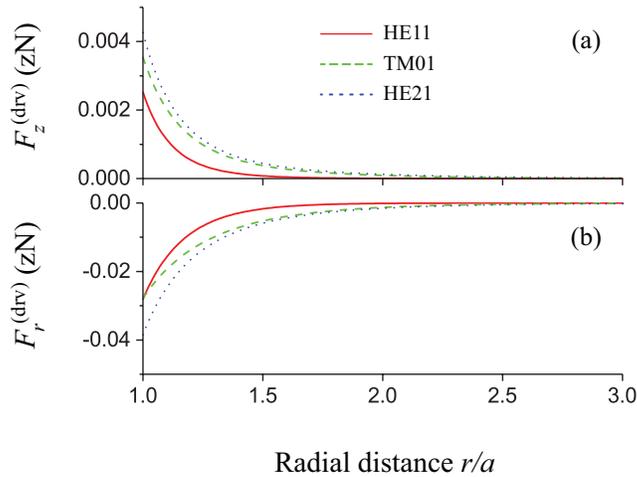}
 \end{center}
\caption{ Radial dependencies of the axial component $F_z^{(\mathrm{drv})}$ (a)
and the radial component $F_r^{(\mathrm{drv})}$ (b) of the driving-field force $\mathbf{F}^{(\mathrm{drv})}$ of a nonresonant guided light field on the atom. The detuning of the field is $\Delta/2\pi=-100$ MHz.
The dipole orientation vector is $\hat{\mathbf{d}}=\hat{\mathbf{x}}$. Other parameters are as for figure \ref{fig2}.}
\label{fig4}
\end{figure}

We plot in figure \ref{fig4} the radial dependencies of the axial component $F_z^{(\mathrm{drv})}$
and the radial component $F_r^{(\mathrm{drv})}$ of the driving-field force in the case where the detuning of the driving field is $\Delta/2\pi=-100$ MHz. The figure shows that, when the detuning $\Delta$ is large, the radial component $F_r^{(\mathrm{drv})}$ is much larger than the axial component $F_z^{(\mathrm{drv})}$.
For the parameters of figure \ref{fig4}, the force $\mathbf{F}^{(\mathrm{drv})}$ for the TE mode
and the azimuthal component $F_\varphi^{(\mathrm{drv})}$ for all the modes vanish and are therefore not plotted.

%%%%%%%%%%%%%%%%%%%%%%% Figure 4

\begin{figure}[tbh]
\begin{center}
 \includegraphics{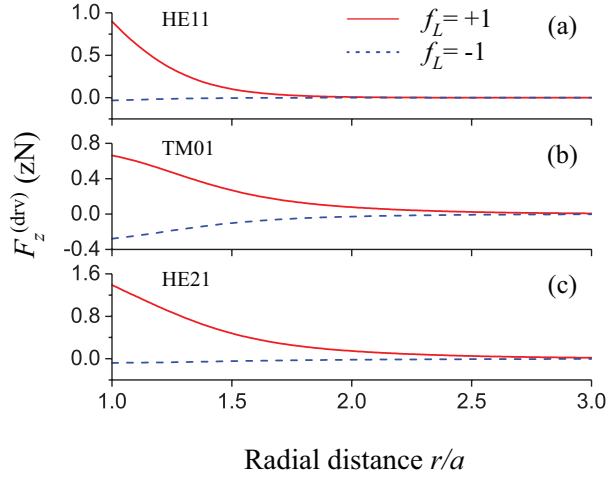}
 \end{center}
\caption{ 
Radial dependence of the axial component $F_z^{(\mathrm{drv})}$ of the driving-field force $\mathbf{F}^{(\mathrm{drv})}$ of a resonant ($\Delta=0$) guided light field on the atom with
a complex dipole matrix element vector $\hat{\mathbf{d}}=(\rmi\hat{\mathbf{x}}-\hat{\mathbf{z}})/\sqrt2$. 
The incident light field is in an $x$-polarized HE$_{11}$ mode (a), a TM$_{01}$ mode (b), or an $x$-polarized HE$_{21}$ mode (c) and propagates in the forward $f_L=+1$ (solid red curves) or backward $f_L=-1$ (dashed blue curves) direction along the fiber axis $z$.  Other parameters are as for figure \ref{fig2}. 
}
\label{fig5}
\end{figure}

Due to the presence of a nonzero axial component $\mathcal{E}_z\propto f_Le_z$ of the guided probe field, the absolute value of the axial component $F_z^{(\mathrm{drv})}$ of the driving-field force may depend on the field propagation direction $f_L$ \cite{chiralforce}.
We plot in figure \ref{fig5} the radial dependence of $F_z^{(\mathrm{drv})}$ in the case where the driving field is at exact resonance with the atom and the dipole orientation vector is $\hat{\mathbf{d}}=(\rmi\hat{\mathbf{x}}-\hat{\mathbf{z}})/\sqrt2$.
We observe from the figure that the absolute value of $F_z^{(\mathrm{drv})}$ depends on $f_L$.
For the parameters of figure \ref{fig5}, the force $\mathbf{F}^{(\mathrm{drv})}$ for the TE mode and 
the components $F_r^{(\mathrm{drv})}$ and $F_\varphi^{(\mathrm{drv})}$ for all the modes vanish and are therefore not plotted.

%%%%%%%%%%%%%%%%%%%%%%% Figure 5

In order to get insight into the origin of the dependence of the driving-field force on the propagation direction, we perform a simple analysis.
For an $x$-polarized hybrid mode or a TM mode with the propagation direction $f_L$, the field at the position of the atom is
$\boldsymbol{\mathcal{E}}(\varphi=0)\propto e_r\hat{\mathbf{x}}+f_Le_z\hat{\mathbf{z}}$. 
For $\mathbf{d}\propto \rmi\hat{\mathbf{x}}-\hat{\mathbf{z}}$, the Rabi frequency is $\Omega\propto \rmi e_r-f_Le_z$.
Since the relative phase between the complex amplitudes $e_r$ and $e_z$ is $\pi/2$, 
the magnitude of $\Omega$ is proportional to $|e_r|-|e_z|$ or $|e_r|+|e_z|$ depending on $f_L$. 
The direction dependence of $\Omega$ leads to the direction dependence of the excited-state population $\rho_{ee}$, which is proportional to $|\Omega|^2$  in the non-saturation regime.
The corresponding difference between the excited-state populations $\rho_{ee}^{(+)}$ and $\rho_{ee}^{(-)}$ for the opposite propagation directions $f_L=+$ and $f_L=-$ is $\Delta\rho_{ee}\equiv\rho_{ee}^{(+)}-\rho_{ee}^{(-)}\propto |e_r||e_z|$. 
This difference is proportional to the electric transverse spin density $\rho_y^{\mathrm{e-spin}}\propto 
\mathrm{Im}[\boldsymbol{\mathcal{E}}^*\times\boldsymbol{\mathcal{E}}]\cdot\hat{\mathbf{y}}
\propto f_L|e_r||e_z|$ of the driving field \cite{highorder}. Due to spin-orbit coupling of light  \cite{Zeldovich,Bliokh review,Bliokh2014,Bliokh review2015,Bliokh2015,Banzer review2015,Lodahl2017}, the sign of $\rho_y^{\mathrm{e-spin}}$ depends on $f_L$. 
The direction dependence of $\rho_{ee}$ leads to that of the absolute value of the force component $F_z^{(\mathrm{drv})}=f_L\hbar \beta_L\Gamma\rho_{ee}$.
Thus, the dependence of $|F_z^{(\mathrm{drv})}|$ on $f_L$ is a signature of spin-orbit coupling of light. 

%%%%%%%%%%%%%%%%%%%%%%% Figure 6
\begin{figure}[tbh]
\begin{center}
 \includegraphics{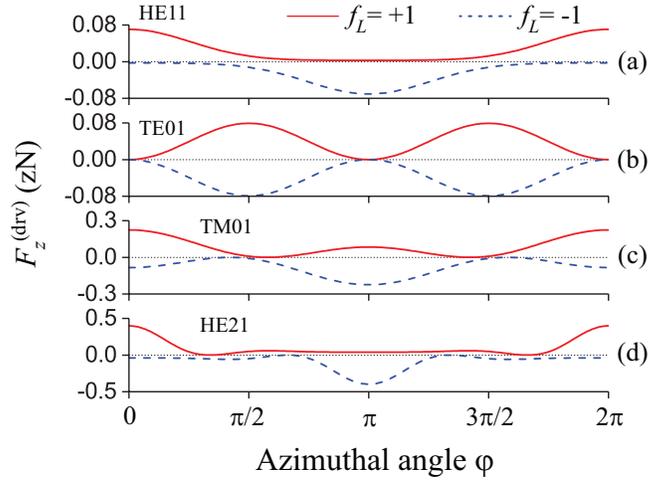}
 \end{center}
\caption{ 
Azimuthal dependence of the axial component $F_z^{(\mathrm{drv})}$ of the driving-field force $\mathbf{F}^{(\mathrm{drv})}$ of a resonant ($\Delta=0$) guided light field on the atom with
a complex dipole matrix element vector $\hat{\mathbf{d}}=(\rmi\hat{\mathbf{x}}-\hat{\mathbf{z}})/\sqrt2$. 
The incident light field is in an $x$-polarized HE$_{11}$ mode (a), a TE$_{01}$ mode (b), a TM$_{01}$ mode (c), or an $x$-polarized HE$_{21}$ mode (d) and propagates in the forward $f_L=+1$ (solid red curves) or backward $f_L=-1$ (dashed blue curves) direction along the fiber axis $z$.  
The distance from the atom to the fiber surface is $r-a=200$ nm.
Other parameters are as for figure \ref{fig2}. 
}
\label{fig6}
\end{figure}

%%%%%%%%%%%%%%%%%%%%%%% Figure 7
\begin{figure}[tbh]
\begin{center}
 \includegraphics{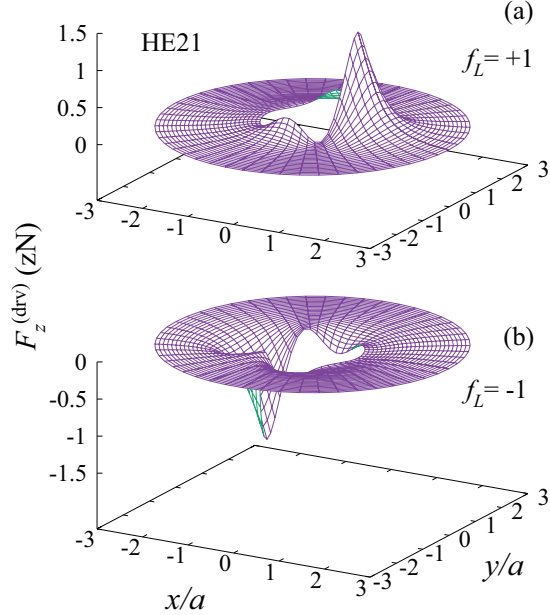}
 \end{center}
\caption{ 
Two-dimensional spatial profile of the axial component $F_z^{(\mathrm{drv})}$ of the driving-field force $\mathbf{F}^{(\mathrm{drv})}$ of a resonant ($\Delta=0$) guided light field on the atom with a complex dipole matrix element vector $\hat{\mathbf{d}}=(\rmi\hat{\mathbf{x}}-\hat{\mathbf{z}})/\sqrt2$. 
The incident light field is in  an $x$-polarized HE$_{21}$ mode and propagates in the forward $f_L=+1$ (a) or backward $f_L=-1$ (b) direction along the fiber axis $z$. Other parameters are as for figure \ref{fig2}. 
}
\label{fig7}
\end{figure}

In general, the driving-field force $\mathbf{F}^{(\mathrm{drv})}$ depends on the azimuthal position $\varphi$ of the atom. We plot in figure \ref{fig6} the azimuthal dependence of the axial component $F_z^{(\mathrm{drv})}$ in the case where the driving field is at exact resonance with the atom and the dipole orientation vector is $\hat{\mathbf{d}}=(\rmi\hat{\mathbf{x}}-\hat{\mathbf{z}})/\sqrt2$. 
In order to get a broader view, we plot in figure \ref{fig7} the spatial profile of $F_z^{(\mathrm{drv})}$ in the fiber transverse plane for an $x$-polarized HE$_{21}$ mode. 
We observe from figures \ref{fig6} and \ref{fig7} that $F_z^{(\mathrm{drv})}$ varies with varying $\varphi$ and depends on $f_L$. For the parameters of figures \ref{fig6} and \ref{fig7}, the components $F_r^{(\mathrm{drv})}$ and $F_\varphi^{(\mathrm{drv})}$ vanish and are therefore not shown.

\subsection{Spontaneous-emission recoil force}

\begin{figure}[tbh]
\begin{center}
 \includegraphics{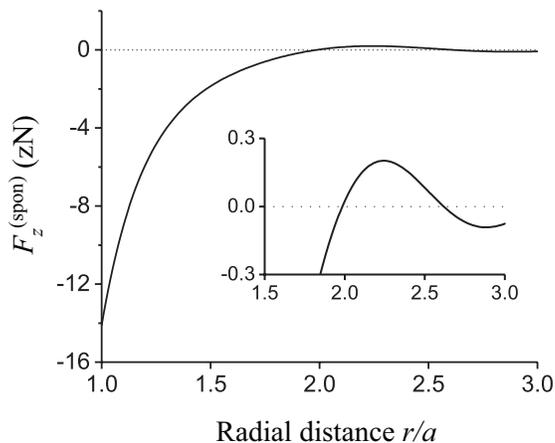}
 \end{center}
\caption{ 
Radial dependence of the axial component $F_{z}^{(\mathrm{spon})}$ of the spontaneous-emission recoil force $\mathbf{F}^{(\mathrm{spon})}$ on the atom with a complex dipole matrix element $\hat{\mathbf{d}}=(\rmi\hat{\mathbf{x}}-\hat{\mathbf{z}})/\sqrt2$. Other parameters are as for figure \ref{fig2}. The inset shows the details of the oscillations of $F_{z}^{(\mathrm{spon})}$
with increasing $r$. 
}
\label{fig8}
\end{figure}

In this subsection, we study the spontaneous-emission recoil force $\mathbf{F}^{(\mathrm{spon})}$.
This force appears when the atomic dipole rotates in the meridional plane containing the atomic position, that is, when
the dipole orientation vector $\hat{\mathbf{d}}$ is a complex vector in the $zx$ plane \cite{Scheel2015,chiralforce}.
We plot in figure \ref{fig8} the radial dependence of the axial component $F_z^{(\mathrm{spon})}$ of the spontaneous-emission recoil force in the case where the dipole orientation vector is $\hat{\mathbf{d}}=(\rmi\hat{\mathbf{x}}-\hat{\mathbf{z}})/\sqrt2$. The figure and its inset show that $F_z^{(\mathrm{spon})}$ oscillates with increasing $r$ and can be negative and positive, depending on the radial position $r$ of the atom \cite{Scheel2015,chiralforce}. The oscillations of $F_z^{(\mathrm{spon})}$ with varying $r$ are due to the oscillations of the decay rate into radiation modes \cite{sponhigh}. Such oscillations result from the interference due to reflections from the fiber surface.
 
%%%%%%%%%%%%%%%%%%%%%%% Figure 8

\begin{figure}[tbh]
\begin{center}
 \includegraphics{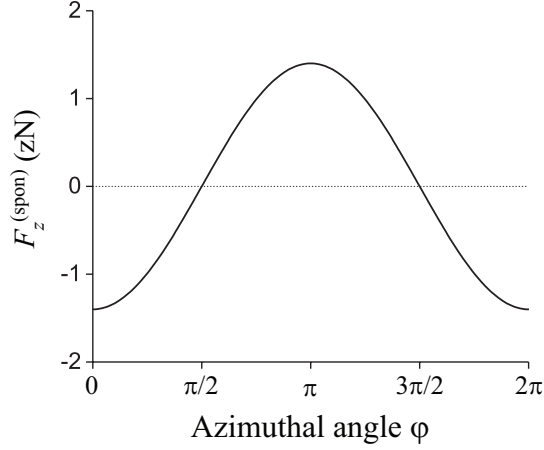}
 \end{center}
\caption{ 
Azimuthal dependence of the axial component $F_{z}^{(\mathrm{spon})}$ of the spontaneous-emission recoil force $\mathbf{F}^{(\mathrm{spon})}$ on the atom with a complex dipole matrix element $\hat{\mathbf{d}}=(\rmi\hat{\mathbf{x}}-\hat{\mathbf{z}})/\sqrt2$. 
The distance from the atom to the fiber surface is $r-a=200$ nm.
Other parameters are as for figure \ref{fig2}. 
}
\label{fig9}
\end{figure}

\begin{figure}[tbh]
\begin{center}
 \includegraphics{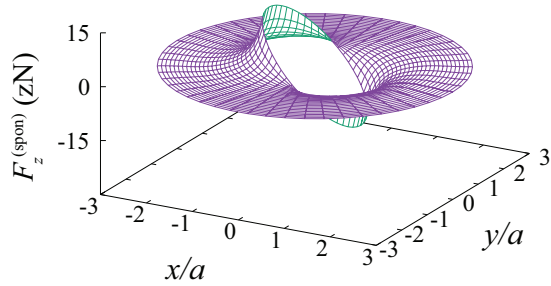}
 \end{center}
\caption{ 
Two-dimensional spatial profile of the axial component $F_{z}^{(\mathrm{spon})}$ of the spontaneous-emission recoil force $\mathbf{F}^{(\mathrm{spon})}$ on the atom with a complex dipole matrix element $\hat{\mathbf{d}}=(\rmi\hat{\mathbf{x}}-\hat{\mathbf{z}})/\sqrt2$. 
Other parameters are as for figure \ref{fig2}. 
}
\label{fig10}
\end{figure}

In general, the spontaneous-emission recoil force $\mathbf{F}^{(\mathrm{spon})}$ depends on the azimuthal position $\varphi$ of the atom. We plot in figure \ref{fig9} the azimuthal dependence of the axial component $F_z^{(\mathrm{spon})}$ in the case where the dipole orientation vector is $\hat{\mathbf{d}}=(\rmi\hat{\mathbf{x}}-\hat{\mathbf{z}})/\sqrt2$. The corresponding spatial profile of $F_z^{(\mathrm{spon})}$ in the fiber transverse plane is shown in figure \ref{fig10}. 
The figures show that the magnitude of $F_z^{(\mathrm{spon})}$ varies with varying $\varphi$.
We also observe that the sign of $F_z^{(\mathrm{spon})}$ depends on $\varphi$. 

%%%%%%%%%%%%%%%%%%%%%%% Figure 9

%%%%%%%%%%%%%%%%%%%%%%% Figure 10

\begin{figure}[tbh]
\begin{center}
 \includegraphics{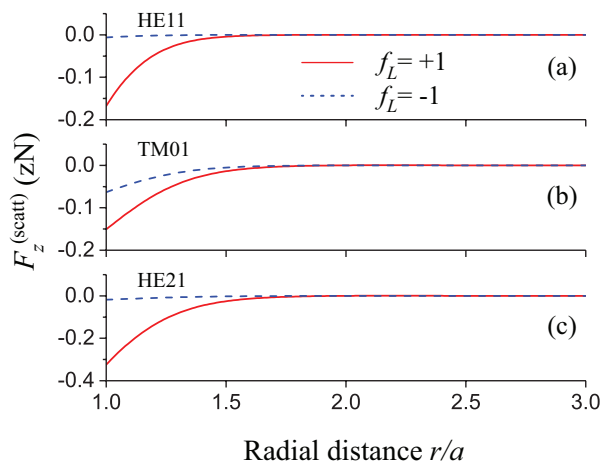}
 \end{center}
\caption{ 
Radial dependence of the axial component $F_{z}^{(\mathrm{scatt})}$ of the scattering recoil force 
$\mathbf{F}^{(\mathrm{scatt})}$ of a resonant ($\Delta=0$) guided light field on the atom with
a complex dipole matrix element vector $\hat{\mathbf{d}}=(\rmi\hat{\mathbf{x}}-\hat{\mathbf{z}})/\sqrt2$. 
The incident light field is in an $x$-polarized HE$_{11}$ mode (a), a TM$_{01}$ mode (b), or an $x$-polarized HE$_{21}$ mode (c) and propagates in the forward $f_L=+1$ (solid red curves) or backward $f_L=-1$ (dashed blue curves) direction along the fiber axis $z$.  Other parameters are as for figure \ref{fig2}.  
}
\label{fig11}
\end{figure}

According to equation (\ref{c25}), the spontaneous-emission recoil force $\mathbf{F}^{(\mathrm{spon})}$ enters the expression for the total force $\mathbf{F}$ with the weight factor $\rho_{ee}$. It is clear that the force produced by the recoil of the scattered photons is
$\mathbf{F}^{(\mathrm{scatt})}=\rho_{ee}\mathbf{F}^{(\mathrm{spon})}$.
We depict in figure \ref{fig11} the radial dependence of the axial component $F_{z}^{(\mathrm{scatt})}$ of the scattering recoil force for the parameters of figure \ref{fig5}.
The figure shows that $F_{z}^{(\mathrm{scatt})}$ depends on the propagation direction $f_L$ of the driving field.
The propagation direction dependence of $F_{z}^{(\mathrm{scatt})}$ results from the propagation direction dependence of $\Omega$.
Due to the evanescent-wave behavior of the radial dependence of the driving field intensity, the oscillations and changes in sign of 
$F_{z}^{(\mathrm{scatt})}$ are small and, hence, hard to see in figure \ref{fig11}. 

%%%%%%%%%%%%%%%%%%%%%%% Figure 11

\begin{figure}[tbh]
\begin{center}
 \includegraphics{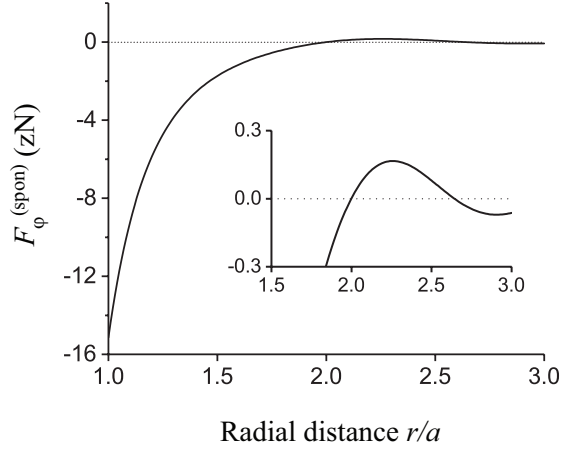}
 \end{center}
\caption{ 
Radial dependence of the azimuthal component $F_{\varphi}^{(\mathrm{spon})}$ 
of the spontaneous-emission recoil force $\mathbf{F}^{(\mathrm{spon})}$ on the atom. 
The dipole matrix element of the atom is $\hat{\mathbf{d}}=(\hat{\mathbf{x}}+\rmi\hat{\mathbf{y}})/\sqrt2$. Other parameters are as for figure \ref{fig2}. The inset shows the details of the oscillations of $F_{\varphi}^{(\mathrm{spon})}$
with increasing $r$. 
}
\label{fig12}
\end{figure}

Note that, due to the cylindrical symmetry, the radial components $F_r^{(\mathrm{spon})}$ and $F_r^{(\mathrm{scatt})}$ of the spontaneous-emission and scattering recoil forces are zero.
Therefore, these force components are not plotted. For the parameters of figures \ref{fig8}--\ref{fig11}, the azimuthal components $F_\varphi^{(\mathrm{spon})}$ and $F_\varphi^{(\mathrm{scatt})}$ vanish and are therefore not shown. However, $F_\varphi^{(\mathrm{spon})}$ and, hence, $F_\varphi^{(\mathrm{scatt})}$ may arise in the case where the dipole orientation vector is a complex vector in the fiber transverse plane $xy$. We plot in figure \ref{fig12} the radial dependence of the azimuthal component $F_\varphi^{(\mathrm{spon})}$ of the spontaneous-emission recoil force in the case where the dipole orientation vector is $\hat{\mathbf{d}}=(\hat{\mathbf{x}}+\rmi\hat{\mathbf{y}})/\sqrt2$. 
Figure \ref{fig12} and its inset show that, like $F_z^{(\mathrm{spon})}$ in the case of figure \ref{fig8}, $F_\varphi^{(\mathrm{spon})}$ oscillates with increasing $r$ and can be negative and positive. For the parameters of figure \ref{fig12}, the axial component $F_z^{(\mathrm{spon})}$ vanishes and is therefore not plotted.

%%%%%%%%%%%%%%%%%%%%%%% Figure 12

\subsection{Fiber-induced van der Waals potential and force}

\begin{figure}[tbh]
\begin{center}
 \includegraphics{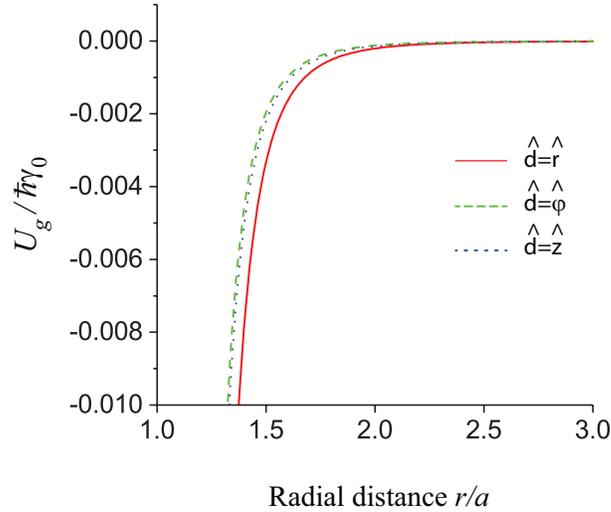}
 \end{center}
\caption{ 
 Radial dependence of the potential $U_g$ for a two-level atom in the ground state. 
The dipole orientation vector is $\hat{\mathbf{d}}=\hat{\mathbf{r}}$ (red solid curve), $\hat{\boldsymbol{\varphi}}$  (green dashed curve), and $\hat{\mathbf{z}}$ (blue dotted curve). Other parameters are as for figure \ref{fig2}. 
}
\label{fig13}
\end{figure}

\begin{figure}[tbh]
\begin{center}
 \includegraphics{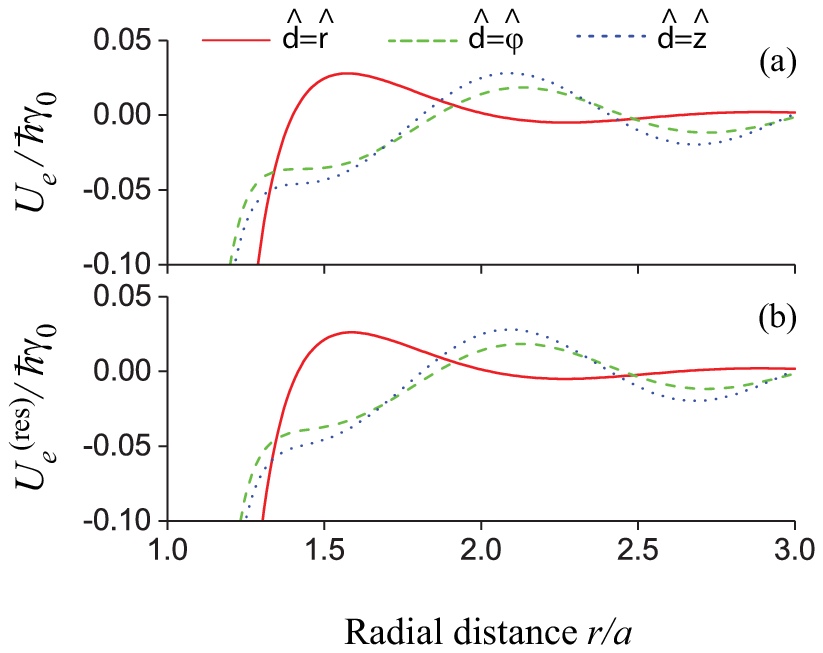}
 \end{center}
\caption{ 
 Radial dependencies of the potential $U_e$ (a) and its resonant part $U_e^{(\mathrm{res})}$ (b) for a two-level atom in the excited state. 
The dipole orientation vector is $\hat{\mathbf{d}}=\hat{\mathbf{r}}$ (red solid curve), $\hat{\boldsymbol{\varphi}}$  (green dashed curve), and $\hat{\mathbf{z}}$ (blue dotted curve). Other parameters are as for figure \ref{fig2}. 
}
\label{fig14}
\end{figure}

In this subsection, we calculate the fiber-induced van der Waals potentials $U_g$ and $U_e$ for the atom in the ground and excited states.
We plot in figures \ref{fig13} and \ref{fig14}(a) the radial dependencies of the potentials $U_g$ and $U_e$, respectively.
We show in figure \ref{fig14}(b) the resonant part $U_e^{(\mathrm{res})}$ of the potential $U_e$ for the excited state.
We observe from the figures that both $U_g$ and $U_e$ depend on the orientation of the atomic dipole.
We also observe that $U_g$ varies monotonically while $U_e$ oscillates with increasing $r$.
The magnitude of $U_e$ is substantially larger than that of $U_g$. 
We recall that the off-resonant part of $U_e$ is $U_e^{(\mathrm{off})}=-U_g$.
Comparison between figures \ref{fig14}(a) and \ref{fig14}(b) shows that  
the potential $U_e$ is mainly determined by its resonant part $U_e^{(\mathrm{res})}$.

%%%%%%%%%%%%%%%%%%%%%%% Figure 13

%%%%%%%%%%%%%%%%%%%%%%% Figure 14

\begin{figure}[tbh]
\begin{center}
 \includegraphics{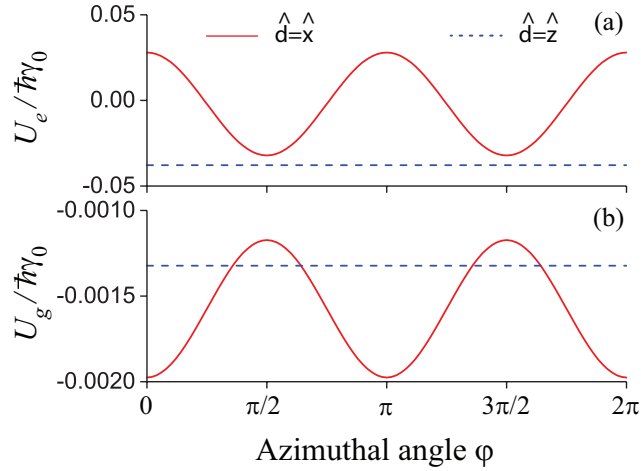}
 \end{center}
\caption{ 
 Azimuthal dependencies of the potentials $U_e$ (a) and $U_g$ (b) for a two-level atom in the excited and ground states. 
The dipole orientation vector is $\hat{\mathbf{d}}=\hat{\mathbf{x}}$ (red solid curve) and $\hat{\mathbf{z}}$  (blue dashed curve). 
The distance from the atom to the fiber surface is $r-a=200$ nm.
Other parameters are as for figure \ref{fig2}. 
}
\label{fig15}
\end{figure}

\begin{figure}[tbh]
\begin{center}
 \includegraphics{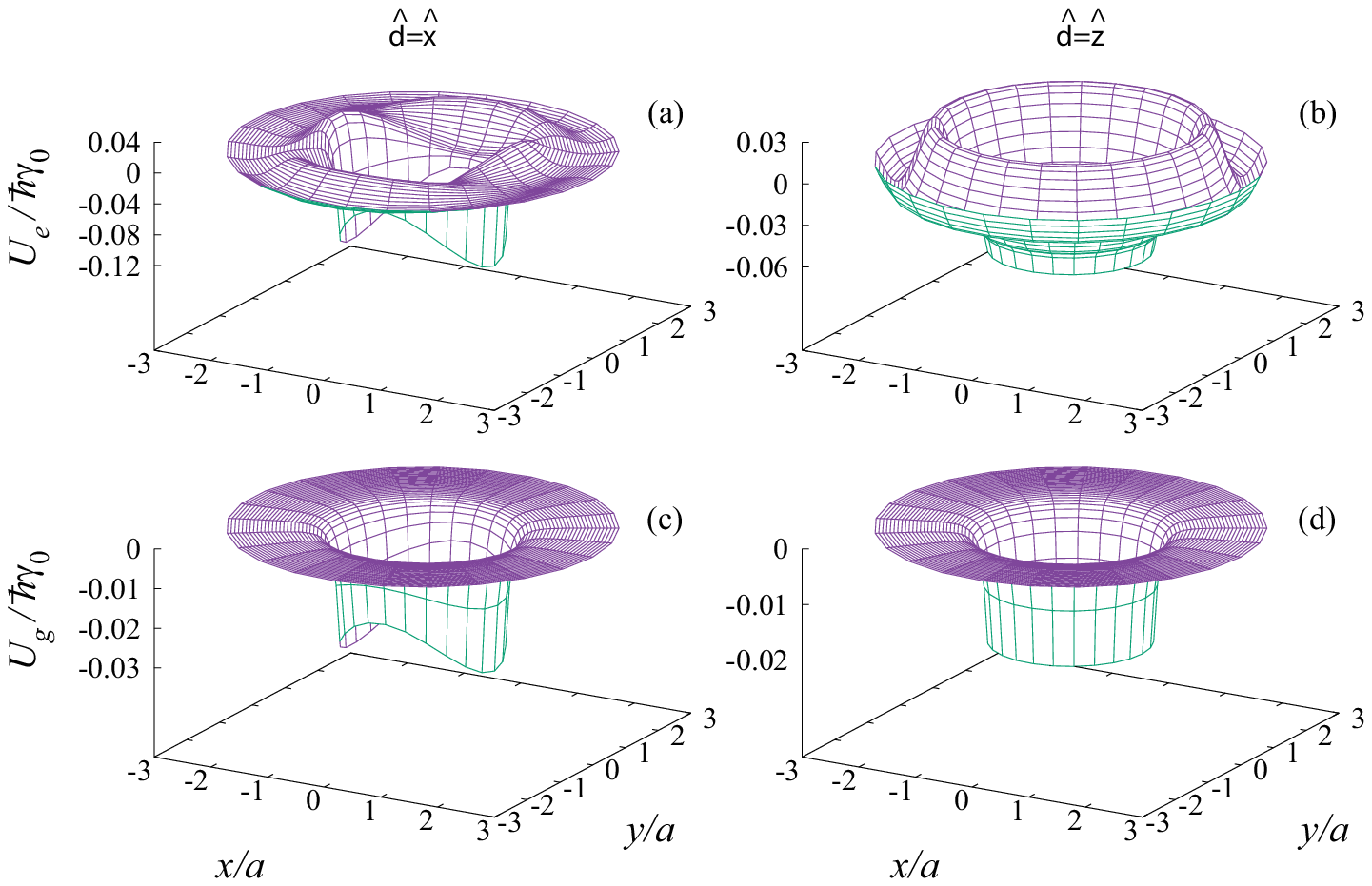}
 \end{center}
\caption{ 
 Two-dimensional spatial profiles of the potentials $U_e$ (upper row) and $U_g$ (lower row) for a two-level atom in the excited and ground states. 
The dipole orientation vector is $\hat{\mathbf{d}}=\hat{\mathbf{x}}$ (left column) and $\hat{\mathbf{z}}$  (right column). 
Other parameters are as for figure \ref{fig2}. 
}
\label{fig16}
\end{figure}

Depending on the dipole orientation, the potentials $U_e$ and $U_g$ may vary with varying azimuthal angle $\varphi$ of the position of the atom in the fiber transverse plane. 
We plot in figure \ref{fig15} the azimuthal dependencies of the potentials in the cases where the dipole orientation vector
is $\hat{\mathbf{d}}=\hat{\mathbf{x}}$ and $\hat{\mathbf{z}}$. The corresponding spatial profiles of the potentials in the fiber transverse plane are shown in figure \ref{fig16}. 
We note that the dependencies of $U_e$ and $U_g$ on $\varphi$ lead to the azimuthal components 
$F_\varphi^{(\mathrm{vdW})e}$ and $F_\varphi^{(\mathrm{vdW})g}$ of the van der Waals forces  (see figure \ref{fig18}).

%%%%%%%%%%%%%%%%%%%%%%% Figure 15

%%%%%%%%%%%%%%%%%%%%%%% Figure 16

\begin{figure}[tbh]
\begin{center}
 \includegraphics{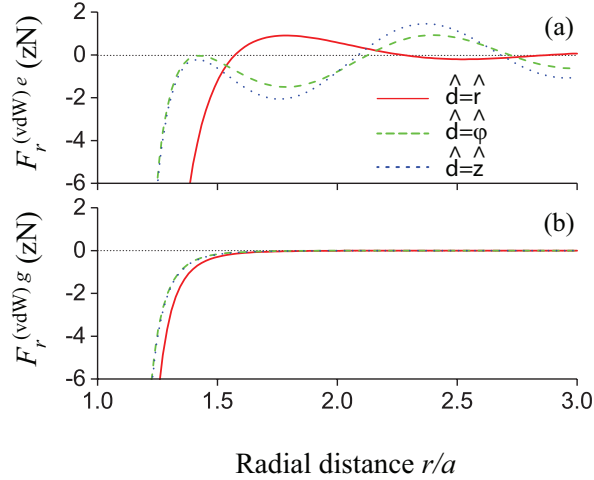}
 \end{center}
\caption{ 
 Radial dependencies of the radial components $F_r^{(\mathrm{vdW})e}$ (a)  and $F_r^{(\mathrm{vdW})g}$ (b)
of the van der Waals forces on the atom in the excited and ground states. 
The dipole orientation vector is $\hat{\mathbf{d}}=\hat{\mathbf{r}}$ (red solid curve), $\hat{\boldsymbol{\varphi}}$  (green dashed curve), and $\hat{\mathbf{z}}$ (blue dotted curve). Other parameters are as for figure \ref{fig2}. 
}
\label{fig17}
\end{figure}

\begin{figure}[tbh]
\begin{center}
 \includegraphics{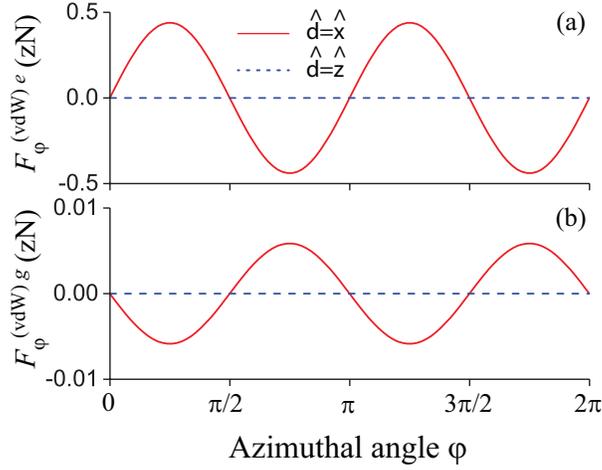}
 \end{center}
\caption{
 Azimuthal dependencies of the azimuthal components $F_\varphi^{(\mathrm{vdW})e}$ (a)  and $F_\varphi^{(\mathrm{vdW})g}$ (b)
of the van der Waals forces on the atom in the excited and ground states.  
The dipole orientation vector is $\hat{\mathbf{d}}=\hat{\mathbf{x}}$ (red solid curve) and $\hat{\mathbf{z}}$  (blue dashed curve). 
The distance from the atom to the fiber surface is $r-a=200$ nm.
Other parameters are as for figure \ref{fig2}. 
}
\label{fig18}
\end{figure}

We show in figure \ref{fig17} the radial dependencies of the radial components $F_r^{(\mathrm{vdW})e}$ and $F_r^{(\mathrm{vdW})g}$
of the van der Waals forces on the atom in the excited and ground states. 
We observe from figure \ref{fig17}(b) that the force $F_r^{(\mathrm{vdW})g}$ for the ground state is always negative and the absolute value of this force reduces  with increasing $r$. Meanwhile, figure \ref{fig17}(a) shows that
the force $F_r^{(\mathrm{vdW})e}$ for the excited state oscillates with increasing $r$, 
and can take not only negative but also positive values depending on the distance $r$.

%%%%%%%%%%%%%%%%%%%%%%% Figure 17

Depending on the orientation of the dipole matrix-element vector $\mathbf{d}$ and the position $(r,\varphi)$ of the atom in the fiber transverse plane,
the azimuthal components $F_\varphi^{(\mathrm{vdW})e}$ and $F_\varphi^{(\mathrm{vdW})g}$
of the van der Waals forces may be nonzero.
We show in figure \ref{fig18} the dependencies of the azimuthal components $F_\varphi^{(\mathrm{vdW})e}$ and $F_\varphi^{(\mathrm{vdW})g}$
of the van der Waals forces on the azimuthal angle $\varphi$ of the position of the atom. 
We observe that $F_\varphi^{(\mathrm{vdW})e}$ and $F_\varphi^{(\mathrm{vdW})g}$ appear when the azimuthal angle between 
the dipole vector $\mathbf{d}$ and the radial vector $\mathbf{r}$ of the atomic position is $\varphi\not=n\pi/2$, with $n$ being an integer number.
In addition, $|F_\varphi^{(\mathrm{vdW})e}|$ and $|F_\varphi^{(\mathrm{vdW})g}|$ achieve their largest values when
$\varphi=\pi/4+n\pi/2$.

%%%%%%%%%%%%%%%%%%%%%%% Figure 18

We note that, in figures \ref{fig13}, \ref{fig14}, \ref{fig16}, \ref{fig17}, the van der Waals potentials and  
the corresponding forces are divergent when $r/a$ is 1. This divergence is a consequence of the fact that, when the distance $r-a$
from the atom to the fiber surface is very small, the van der Waals potential of the atom near the fiber can be approximated by the van der Waals potential of an atom near a flat dielectric surface, which is proportional to $-1/(r-a)^3$.

\subsection{Total force}

\begin{figure}[tbh]
\begin{center}
 \includegraphics{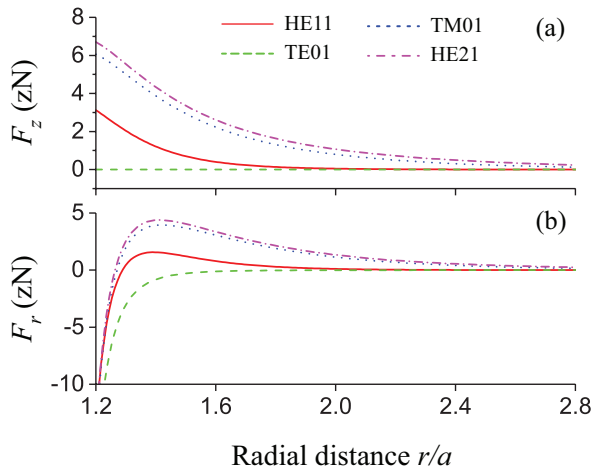}
 \end{center}
\caption{ Radial dependencies of the axial component $F_z$ (a) and the radial component $F_r$ (b) of the total force $\mathbf{F}$ of the field
on the atom with the dipole orientation vector $\hat{\mathbf{d}}=\hat{\mathbf{x}}$.
The detuning of the driving field from the atom in the absence of the fiber is $\Delta_0/2\pi=10$ MHz. 
The power of the driving field is $P=100$ pW. The fiber-induced shift of the atomic transition frequency is taken into account.
Other parameters are as for figure \ref{fig2}.}
\label{fig19}
\end{figure}

Finally, we compute the total force $\mathbf{F}$ of the field on the atom.
We plot in figure \ref{fig19} the radial dependencies of the axial component $F_z$ and the radial component $F_r$ of the total force $\mathbf{F}$ of the field
on the atom in the case where the dipole orientation vector is $\hat{\mathbf{d}}=\hat{\mathbf{x}}$. In these numerical calculations, we take into account 
the effect of the fiber-induced van der Waals potentials on the detuning of the driving field from the atomic transition frequency.
Since $\mathbf{d}$ is a real vector, we have $\mathbf{F}^{(\mathrm{spon})}=0$ and, hence, $F_z=F_z^{(\mathrm{drv})}$. 
The radial component $F_r$ of the total force is composed of the radial component $F_r^{\mathrm{(drv)}}$ of the driving-field force and 
the radial components $F_r^{\mathrm{(vdW)}e}$ and $F_r^{\mathrm{(vdW)}g}$ of the fiber-induced van der Waals forces with the weight factors
$\rho_{ee}$ and $\rho_{gg}$, respectively.
For the parameters of figure \ref{fig19}, the azimuthal component $F_\varphi$ of the total force is zero and is therefore not shown. 
We observe from figure \ref{fig19}(a) that the axial component $F_z$ of the force of the HE$_{21}$ mode is larger than that of the other modes. Figure \ref{fig19}(b) shows that the radial component $F_r$ of the total force can be positive or negative depending on the position $r$ of the atom. The repulsive feature of the force in the region of large $r$ is mainly due to the fact that a positive detuning $\Delta_0$ was used in the calculations.

%%%%%%%%%%%%%%%%%%%%%%% Figure 19

\begin{figure}[tbh]
\begin{center}
 \includegraphics{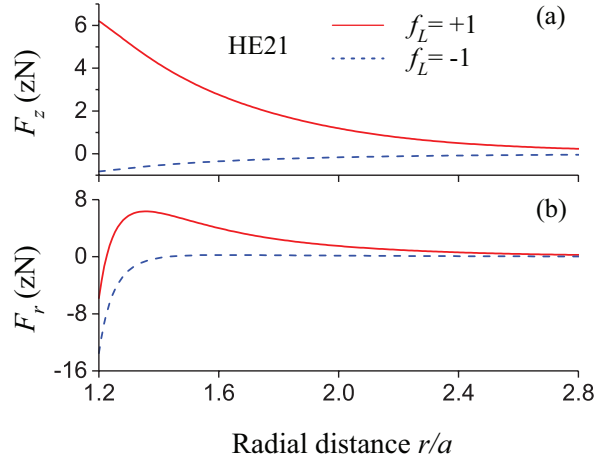}
 \end{center}
\caption{ 
 Radial dependencies of the axial component $F_z$ (a) and the radial component $F_r$ (b) of the total force $\mathbf{F}$ of the field
on the atom with the complex dipole orientation vector $\hat{\mathbf{d}}=(\rmi\hat{\mathbf{x}}-\hat{\mathbf{z}})/\sqrt2$. 
The incident light field is in an $x$-polarized HE$_{21}$ mode and propagates in the forward $f_L=+1$ (solid red curves) 
or backward $f_L=-1$ (dashed blue curves) direction along the fiber axis $z$. 
The detuning of the driving field from the atom in the absence of the fiber is $\Delta_0/2\pi=10$ MHz. 
The power of the driving field is $P=100$ pW. The fiber-induced shift of the atomic transition frequency is taken into account.
Other parameters are as for figure \ref{fig2}. 
}
\label{fig20}
\end{figure}

\begin{figure}[tbh]
\begin{center}
 \includegraphics{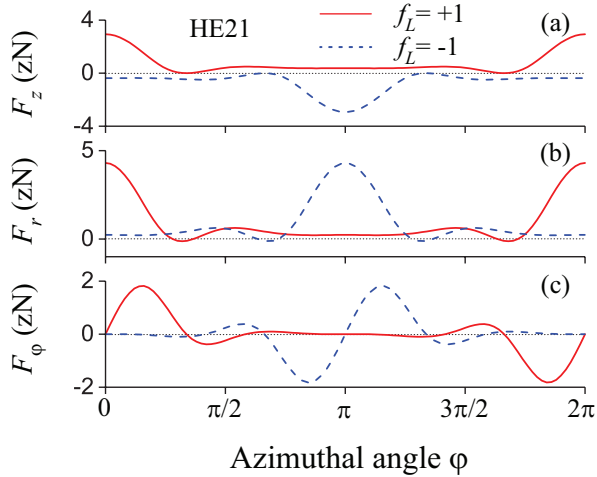}
 \end{center}
\caption{ 
 Azimuthal dependencies of the axial component $F_z$ (a), the radial component $F_r$ (b), and the azimuthal component $F_\varphi$ (c) of the total force $\mathbf{F}$ of the field
on the atom with the complex dipole orientation vector $\hat{\mathbf{d}}=(\rmi\hat{\mathbf{x}}-\hat{\mathbf{z}})/\sqrt2$. 
The distance from the atom to the fiber surface is $r-a=200$ nm.
Other parameters are as for figure \ref{fig20}. 
}
\label{fig21}
\end{figure}

When the atomic dipole matrix-element vector is a complex vector, the propagation direction dependence of the Rabi frequency and the asymmetric spontaneous emission may occur as shown earlier. In this case, the absolute value of the total force may depend on the propagation direction of the probe field. 
We plot in figures \ref{fig20} and \ref{fig21} the radial and azimuthal dependencies of the components of the total force $\mathbf{F}$ in the case where the dipole orientation vector is a complex vector $\hat{\mathbf{d}}=(\rmi\hat{\mathbf{x}}-\hat{\mathbf{z}})/\sqrt2$.
The incident light field is in an $x$-polarized HE$_{21}$ mode.
The figure shows that the absolute values of components of the force depend on the propagation direction $f_L$.
For the parameters of figure \ref{fig20}, where $\varphi=0$, the azimuthal component $F_\varphi$ vanishes and is therefore not shown in this figure.
However, in the case of figure \ref{fig21}, where $\varphi$ is arbitrary, $F_\varphi$ may become nonzero.

%%%%%%%%%%%%%%%%%%%%%%% Figure 20

%%%%%%%%%%%%%%%%%%%%%%% Figure 21

\section{Summary}
\label{sec:summary}

In this work, we have calculated analytically and numerically the force of light on a two-level atom near an ultrathin optical fiber.
We have derived the expressions for the force in terms of the mode functions and the Green tensor. We have shown that the total force consists of the driving-field force, the spontaneous-emission recoil force, and the fiber-induced van der Waals potential force. The axial component of the driving-field force is a light pressure force, while
the radial component is a gradient force. The azimuthal component of the driving-field force may also appear and is, in general, a combination of the pressure and gradient forces in the azimuthal direction.
Due to the existence of a nonzero axial component of the field in a guided mode, 
the Rabi frequency and hence the magnitude of the force of the guided driving field may depend on the propagation direction.  
When the atomic dipole rotates in the meridional plane, the spontaneous-emission recoil force may arise as a result of the asymmetric spontaneous emission with respect to opposite propagation directions. The spontaneous-emission recoil force has a nonvanishing axial or azimuthal component when the atomic dipole rotates in the meridional or cross-sectional plane, respectively. The van der Waals potential for the atom in the ground state is off-resonant and opposite to the off-resonant part of the van der Waals potential for the atom in the excited state. Unlike the potential for the ground state, the potential for the excited state has a resonant part, which is dominant with respect to the off-resonant part, and may oscillate depending on the distance from the atom to the fiber surface.

Our results are fundamental, as they quantify a new physical behavior of the force of light. They can also be envisioned to have significant influence on ongoing and future experiments in quantum and atom optics. Having a controllable force of a structured light field on atoms can help to develop near-field optics, break the existing limits, and reach new dynamical regimes.

\ack
We acknowledge support for this work from the Okinawa Institute of Science and Technology Graduate University.
D. K. and M. P. acknowledge the supports by the Russian Foundation for Basic Research (projects No 16-32-60167 and No 18-32-00691)
and by the BASIS Foundation.

%%%%%%%%%%%%%%%%%%%%%%%%%%%%%%%%%%%%%%%%%%
%%%%%%%%%%%%%%%%%%%%%%%%%%%%%%%%%%%%%%%%%%

\appendix

\section{Green tensor in terms of mode functions}
\label{sec:connection}

The generalized Green tensor $\mathbf{G}$ can be decomposed as $\mathbf{G}=\mathbf{G}_{\mathrm{gd}}+\mathbf{G}_{\mathrm{rd}}$, where the parts $\mathbf{G}_{\mathrm{gd}}$ and $\mathbf{G}_{\mathrm{rd}}$ are related to guided and radiation modes, respectively. For clarity, we now use the explicit expressions $\mu=(\omega N f p)$ and $\nu=(\omega \beta l p)$ for the mode indices of guided and radiation modes, respectively.
According to  \cite{Tromborg}, the parts $\mathbf{G}_{\mathrm{gd}}$ and $\mathbf{G}_{\mathrm{rd}}$ are given in the upper half-plane of the complex frequency $\omega_c$ as
\begin{eqnarray}\label{green1}
\mathbf{G}_{\mathrm{gd}}(\mathbf{R},\mathbf{R}';\omega_c)&=&
\frac{c^2}{2\pi}\int _0^{\infty}\rmd\omega \sum_{Nfp}\beta'(\omega)
\frac{\mathbf{e}^{(\omega Nfp)}(\mathbf{r})\mathbf{e}^{(\omega Nfp)*}(\mathbf{r}')}{\omega^2-\omega_c^2}
\nonumber\\&&\mbox{}
\times\rme^{\rmi pl(\varphi-\varphi')}\rme^{\rmi f\beta(z-z')},\nonumber\\
\mathbf{G}_{\mathrm{rd}}(\mathbf{R},\mathbf{R}';\omega_c)&=&
\frac{c^2}{2\pi}\int\limits_{0}^{\infty}\rmd\omega\int\limits _{-kn_2}^{kn_2}\rmd\beta\sum_{lp}
\frac{\mathbf{e}^{(\omega\beta lp)}(\mathbf{r})\mathbf{e}^{(\omega\beta lp)*}(\mathbf{r}')}{\omega^2-\omega_c^2}
\nonumber\\&&\mbox{}
\times\rme^{\rmi l(\varphi-\varphi')}
\rme^{\rmi \beta(z-z')},
\end{eqnarray}
where $\mathbf{r}=(r,\varphi)$ and $\mathbf{r}'=(r',\varphi')$. 
We can show that the Green tensor satisfies the Schwarz reflection principle 
$\mathbf{G}^*(\mathbf{R},\mathbf{R}';\omega_c)=\mathbf{G}(\mathbf{R},\mathbf{R}';-\omega_c^*)$ \cite{Raabe}.

On the real $\omega$ axis,
the Green tensor is defined as $\mathbf{G}(\mathbf{R},\mathbf{R}';\omega)=\lim_{\epsilon\to 0+}\mathbf{G}(\mathbf{R},\mathbf{R}';\omega+\rmi\epsilon)$. 
We use the identity 
\begin{equation}\label{green2}
\frac{1}{x-\rmi\epsilon}=\mathcal{P}\frac{1}{x}+\rmi\pi\delta(x),
\end{equation}
where $\epsilon\to0+$. 
Then, for a positive real frequency $\omega_0$, we find
\begin{eqnarray}\label{green3}
\mathrm{Re}\,\mathbf{G}_{\mathrm{gd}}(\mathbf{R},\mathbf{R}';\omega_0)&=&
\frac{c^2}{2\pi}\mathcal{P}\int _0^{\infty}\rmd\omega \sum_{Nfp}\beta'(\omega)
\frac{\mathbf{e}^{(\omega Nfp)}(\mathbf{r})\mathbf{e}^{(\omega Nfp)*}(\mathbf{r}')}{\omega^2-\omega_0^2}
\nonumber\\&&\mbox{}
\times\rme^{\rmi pl(\varphi-\varphi')}\rme^{\rmi f\beta(\omega)(z-z')},\nonumber\\
\mathrm{Re}\,\mathbf{G}_{\mathrm{rd}}(\mathbf{R},\mathbf{R}';\omega_0)&=&
\frac{c^2}{2\pi}\mathcal{P}\int\limits_{0}^{\infty}\rmd\omega\int\limits _{-kn_2}^{kn_2}\rmd\beta\sum_{lp}
\frac{\mathbf{e}^{(\omega\beta lp)}(\mathbf{r})\mathbf{e}^{(\omega\beta lp)*}(\mathbf{r}')}{\omega^2-\omega_0^2}
\nonumber\\&&\mbox{}
\times\rme^{\rmi l(\varphi-\varphi')}\rme^{\rmi \beta(z-z')},
\end{eqnarray}
and
\begin{eqnarray}\label{green4}
\mathrm{Im}\,\mathbf{G}_{\mathrm{gd}}(\mathbf{R},\mathbf{R}';\omega_0)&=&
\frac{c^2}{4\omega_0}\sum_{Nfp}\beta'(\omega_0)\mathbf{e}^{(\omega_0 Nfp)}(\mathbf{r})\mathbf{e}^{(\omega_0 Nfp)*}(\mathbf{r}')
\nonumber\\&&\mbox{}
\times\rme^{\rmi pl(\varphi-\varphi')}\rme^{\rmi f\beta(\omega_0)(z-z')},\nonumber\\
\mathrm{Im}\,\mathbf{G}_{\mathrm{rd}}(\mathbf{R},\mathbf{R}';\omega_0)&=&
\frac{c^2}{4\omega_0}\int\limits _{-k_0n_2}^{k_0n_2}\rmd\beta\sum_{lp}\mathbf{e}^{(\omega_0\beta lp)}(\mathbf{r})\mathbf{e}^{(\omega_0\beta lp)*}(\mathbf{r}')
\nonumber\\&&\mbox{}
\times\rme^{\rmi l(\varphi-\varphi')}\rme^{\rmi \beta(z-z')}.
\end{eqnarray}
In deriving the above equations, we have taken into account the facts that the results of the summations
$\sum_{Nfp}\mathbf{e}^{(\omega Nfp)}(\mathbf{r})\mathbf{e}^{(\omega Nfp)*}(\mathbf{r}')\rme^{\rmi pl(\varphi-\varphi')}\rme^{\rmi f\beta(z-z')}$ 
and $\sum_{lp}\mathbf{e}^{(\omega\beta lp)}(\mathbf{r})\mathbf{e}^{(\omega\beta lp)*}(\mathbf{r}')\rme^{\rmi l(\varphi-\varphi')}\rme^{\rmi \beta(z-z')}$ are real tensors.
When we use the properties of the mode functions, we can show that $G_{r\varphi}(\mathbf{R},\mathbf{R};\omega)=G_{\varphi z}(\mathbf{R},\mathbf{R};\omega)=G_{zr}(\mathbf{R},\mathbf{R};\omega)=0$.

\end{document}